# Electrochemical Evaluation of Mg and a Mg-Al 5%Zn Metal Rich Primers for Protection of Al-Zn-Mg-Cu Alloy in NaCl

A. Korjenic ǂ *, L. Blohm **, J.R. Scully*

Keywords: Aluminum alloy, sacrificial anode, cathodic protection, aluminum primer, x-ray diffraction, electrochemical impedance spectroscopy

* Center for Electrochemical Science and Engineering, Department of Materials Science and Engineering, University of Virginia, Charlottesville, VA 22904

**DEVCOM Army Research Laboratory, Aberdeen Proving Ground, MD 21005, USA

ǂ Corresponding author. E-mail: ak4m@virignia.edu

## Abstract

High purity magnesium and a Mg-Al 5wt% Zn metal rich primer (MRP) were compared for their ability to suppress intergranular corrosion (IGC) and intergranular stress corrosion cracking (IG-SCC) in peak aged AA 7075-T651 by sacrificial anode-based cathodic prevention. Tests were conducted in 0.6 M NaCl solution under full immersion. These evaluations considered the ability of the primer to attain an intermediate negative open circuit potential (OCP) such that the galvanic couple potential with bare aluminum alloy (AA) 7075-T651 resided below a range of potentials where IGC is prevalent. The ability of the primer to achieve an OCP negative enough that the AA 7075-T651 could be protected by sacrificial anode-based cathodic prevention and the ability to sustain this function over time were evaluated as a first step by utilizing a NaCl solution. The primers consisted of epoxy resins embedded with either (1) Mg flake pigments (MgRP) or (2) Mg flake pigments and spherical Al-5 wt.% Zn together as a composite (MgAlRP). MgRP was an effective coating for cathodic protection but dispensed less anodic charge than the composite MgAlRP. Cross-sectional analysis demonstrated that some Mg flakes dissolved while uniform surface oxidation occurred on the remaining Mg flakes which led to impaired activation. The composite MgAlRP maintained a suitably negative OCP over time, remained activated, dispensed high anodic charge, and remained an anode in zero resistance ammeter testing. Chemical stability modeling and zero resistance ammeter testing suggest that Mg corrosion elevates the pH which dissolved aluminum oxides and hydroxide thereby activates the Al-5wt.% Zn pigments, thereby providing a primary (i.e. Mg corrosion) and secondary process to enable superior (activation of Al-5wt%Zn) sacrificial anode-based cathodic protection.

## Introduction

Aircraft aluminum alloys range widely in composition due to a large range in the demand for optimal material properties and performance. The wing-spar of commercial aircraft structures is largely comprised of 7XXX-series aluminum alloys[1–3]. Aluminum alloy (AA) 7075 is a naturally aged precipitation hardened Al-Zn-Mg-Cu alloy that owes its IGC susceptibility to its grain boundary phase $MgZn_2$, known as $\eta$ phase. The 7XXX-series aluminum alloy AA 7075-T651 is commonly used commercial aircraft. This alloy is hardened through peak-aging and stress relieved by stretching (3%). Peak age (T6x) hardening of AA 7075-T651 results in a maximum yield strength (0.2% offset) of 455 MPa[2,4,5]; however, this process is also responsible for the increased susceptibility to intergranular corrosion (IGC) and intergranular stress corrosion cracking (IGSCC)[4,6–10]. The enhanced localized corrosion susceptibility of AA 7075-T651 is due to its heterogeneous microstructure and a wide range of constituent particles and precipitates introducing local chemical inhomogeneity and enhanced localized breakdown in the form of matrix-phase boundary attack[11–16], selective dissolution[17–22], and pitting[11,13,23–26]. The peak aging of AA 7075 results in a greater phase fraction of intragranular precipitates and grain boundary ($\eta$ phase) precipitates as well as solute-depleted zones[4,12,27]. Intragranular coherent precipitates increase the yield strength of AA 7075-T651, but the peak aging treatment also forms heterogeneously nucleated $\eta$ phase increasing IGC/IGSCC susceptibility [1,2,4,5,7,28–31].

Marine environments, which are abundant in $Cl^-$ and subject to wet-dry cycles, are particularly aggressive increasing the susceptibility for environmentally assisted cracking (EAC) and stress corrosion cracking (SCC) [2,32–36]. EAC of 7xxx-series aluminum alloys in both aqueous environments and water vapor have been studied extensively with numerous proposed models aimed at understanding SCC mechanisms [2,6,8,31,37–41]. Localized breakdown contributes to EAC as environmental exposure facilitates the breakdown and evolution of pits on the surface acting as stress concentrators ultimately lowering the SCC resistance[7,31,42,43]. The IG-SCC behavior is hypothesized to be governed by a coupled anodic dissolution process (i.e. electro-dissolution of the grain boundary $\eta$ phase and the surrounding matrix), which also catalyzes the formation of an aggressive acidified local crack tip chemistry that, in turn, facilitates the enhanced generation of crack tip hydrogen and uptake enabling embrittlement in the fracture process zone[10,36,44–52]. The IG-SCC crack initiation tendencies and growth rate of AA 7075-T651 exhibit a great deal of potential dependence under potentiostatic conditions (PS) in 0.6 M NaCl[28,49]. Therefore, the mitigation of both IGC and IG-SCC is possible through the establishment of potentials that are more negative than the pitting potential, $E_{pit}$, of the $\eta$ phase, $E_{pit\,(\eta)}$[39,53,54]. The potential dependence of environmentally induced fracture susceptibility of AA 7075-T651 has been investigated previously by Harris et al., who reported variation in the crack growth rate and threshold stress intensity with applied potential[28]. Stage II crack growth rate develops a minimum potential between -0.9 $V_{SCE}$ and -1.0 $V_{SCE}$. This range of minimum crack growth potentials is theorized to minimize $\eta$ phase corrosion, hydrolysis, and acidification. The environmental fracture susceptibility of numerous 7xxx series aluminum alloys such as AA 7050-T651 exhibits an identical potential window for crack growth suppression[55]. This illuminates a possible mitigation strategy for this class of alloys for use in marine service such that stage II crack growth is minimized based on the establishment of potentials more negative than the pitting potential of the η phase, $E_{pit\,(\eta)}$. This situation can be described by the term "sacrificial cathodic prevention" where the goal is to polarize below selected critical potentials where susceptibility is indicated[25,56,57].

Protecting against environmental fracture whether it be SCC or IGSCC is a larger challenge than providing static protection against localized or uniform corrosion. The challenge is to develop a substrate protection system that can suppress environmental fracture. Traditional legacy systems have used chromate-based coatings, which function based on the release of chromate ions, to chemically inhibit corrosion on aluminum alloys[17,58–65]. Although chromate-based coatings are very effective, there are concerns over the toxicity of hexavalent chromium as well as growing requirements in the aerospace industry that motivate the quest for alternative coating systems[66–71]. Metal rich primers are often accompanied by anodizing treatments to the substrate, conversion coatings, and the application of a topcoat layer[58,60,72,73]. Conversion coatings vary in composition from Cr [59,60,64,74–76], Zr[77–79], $Zn_3(PO_4)_2$[80–83], or Ce[,84–86]. They generally serve the purpose of providing corrosion inhibition through the dissolution of cation species. The MRP layer is found between the anodizing layer and the top-coat layer; this arrangement functions as an active coating system that has multiple modes of protection. In general, an MRP coating system is comprised of a metallic pigment embedded within an epoxy-based resin.

MRP protection of an aluminum alloy substrate is multifaceted and may include 1) sacrificial anode based cathodic protection of the substrate, particularly through intermetallic compounds (IMC) which normally induce micro-galvanic attack with the FCC matrix phase, 2) chemical inhibition of oxygen reduction reaction (ORR) at some IMC, and 3) a secondary barrier effect. The corrosion performance of an MRP is intimately tied to these three possible forms of protection, which vary based on the physical coating attributes such as pigment volume concentration (PVC) and type of pigment used. A range of MRP with varying pigment chemistries (Al, Mg, Zn) has been tested on 2XXX and 5XXX-series aluminum alloys, indicating a capacity to provide active protection of the micro-galvanic couple between the IMC and matrix[54,58,60,72,74,87–102]. However, mitigating environmental fracture remains an unmet challenge. Previous work on zinc-rich primers applied to AA 5456-H116 indicated that IG-SCC mitigation was achieved when using either cold mounted Zn anode or a Zn-based metal rich primer (ZnRP) coating[99,103]. Mitigation was not observed when a topcoat was applied to an inorganic ZnRP[99]. The chemical effects of $Zn^{2+}$ were assessed in a complimentary evaluation separating any cathodic protection effects afforded by the primer and the addition of $ZnCl_2$ salt. In this case, the observed IG-SCC growth rate was reduced by as much as three orders of magnitude over a range of stress intensity in the presence of just $Zn^{2+}$ ions[104].

A few criteria must be satisfied for a given MRP to provide sacrificial anode based cathodic protection to an underlying substrate which include 1) an electrical connection between the AA 7075-T651 substrate and electrolyte, 2) an ionic connection between the AA 7075-T651 substrate and electrolyte, and 3) contain pigment which maintains the ability to support an anodic reaction leaving the substrate to support the cathodic reaction. In this situation a mixed potential based galvanic couple is formed between the pigment within the coating and the AA 7075-T651 substrate and is polarized below the corrosion potential, $E_{corr}$, of the AA 7075-T651 ($E_{Corr,\ 7075}$) [98,99]. Ideally the mixed potential is also capable of suppressing below critical threshold potential such as $E_{pit\,(\eta)}$ [39,89]. There is a lack of prior studies assessing the corrosion characteristics and coating performance of these MRPs applied to 7xxx- series aluminum alloys susceptible to EAC with these target goals in mind. An effective MRP might also provide chemical inhibition via pigment dissolution and redeposition of beneficial species within macro-defects such as scribes/scratches as well as preventing blistering and under-paint corrosion[72,98,105].

In a complementary study, preliminary experiments were conducted assessing the electrochemical behavior of an AlRP applied to AA 7075-T651 consisting of Al-5wt%Zn pigment[89]. This AlRP was not shown to provide potential suppression below the $E_{pit\,(\eta)}$ when tested in 0.6 M NaCl[89]. The electrochemical investigation measured a cathodic current density over the AlRP-coated AA 7075-T651 sample during galvanic coupling with bare AA 7075-T651 which indicated the AlRP coating is operating as a cathode instead of an anode to the AA 7075-T651[89]. Additional electrochemical diagnostic testing assessing the ability of the AlRP coating to discharge anodic current at a potential deemed protective, below the $E_{pit\,(\eta)}$ (-0.85 $V_{SCE}$), concluded that the coating discharges cathodic current opposite to the intended sacrificial anode capability[89]. Examination of the oxidation behavior of the intact buried AlRP during ASTM B117 salt spray exposure indicated no oxidation of the coating-electrolyte interface and the interior of the coating[89]. The combination of measured electrochemical behavior and the negligible oxidation observed suggested the AlRP was inadequate to protect AA 707-T651[89]. It was noted that the alloying of 5wt%Zn in the Al pigment is likely insufficient to provide adequate sacrificial anode-based cathodic protection to AA 7075-T651[89].

MgRPs are known to be effective at providing sacrificial anode-based cathodic prevention, scribe defect protection, and possess high impedance secondary barrier effects on aluminum alloys[90,96,98,106,107]. The use of MgRP's, first developed by Bierwagen et al., was introduced for the protection of aluminum alloys[90,108–111]. The galvanic effects of a MgRP applied to AA 2024-T3 provide sacrificial anode based cathodic protection via cathodic polarization of the substrate and have been shown to provide protection to a defect region sufficient to suppress pitting[72,74,98,105,112]. The more challenging verification of $\eta$ phase protection was not attempted. Polarization of AA 7075-T651 below $E_{pit\ (\eta)}$ might be better achieved with Mg and MgO rich primer due to its low electrode potential[72,87,98]. However, the criterion for protection proposed herein was not attempted elsewhere in the literature. The use of high purity Mg pigment in MgRP is limited by its high self-corrosion rate[58,60,98,102]. The dissolution of Mg pigment to $Mg^{2+}$ within MgRP was shown to precipitate corrosion products within the scribed region, providing a surface modified layer to the otherwise bare defect[60,72,73,87,102,113].

Studies on the influence of MgO or derivative compounds on aluminum alloys have shown similar corrosion performance, as seen with zinc-rich primers for the galvanic protection of steel and their fasteners[72,87,114–119]. These MgO pigments may dissolve and precipitate at the substrate and modify the Al surface by filling pores within the Mg oxide layer, increasing the stability of the layer to $Cl^{-}$[72,87]. Studies performed on AA 2024-T3 in chloride-containing environments showed that the introduction of $Mg^{2+}$ ions leads to a pH rise and a negative shift in the $E_{corr}$ common to aluminum alloys exposed to alkaline environments given its amphoteric nature. The dissolution of Mg from either the coating or the dissolution of Mg-based corrosion products such as $Mg(OH)_2$ enables the supply of a reservoir of $Mg^{2+}$ similar to a coating that dispenses a possible chemical inhibitor. Here, $Mg^{2+}$ products precipitate chemically in a scratch at high pH sites, such as IMCs, where the pH becomes quite alkaline[72,87]. Such repartitioning of $Mg^{2+}$ was observed in the case of MgRP and MgORP on AA 2024-T3[60,72,87,102,112]. These repartitioning effects occur due to chemical dissolution of the passivated MgO pigment (MgORP). A greater amount of $Mg^{2+}$ repartitioned for MgRP than MgORP. However, both were observed to exhibit similar amounts of reduced corrosion damage within the scribe after 2.5 years of field exposure at Kennedy Space Center (KSC)[72,87,98,100]. The work of Mokaddem et al. shows the variation in the dissolution rate of Mg, Al, and 2024 (Al-Cu-Mg) using atomic emission spectroscopy, indicating that when Mg is present in the solid solution of 2024, the co-dissolution of Mg and Al occurs together until the pH increases, allowing the Mg to form $Mg(OH)_2$ and thereby reducing the dissolution of Mg and increasing that of Al[120]. In Mg and MgO pigments, the MgO species was soluble at the initial neutral pH solutions containing $Cl^-$[100,102]. It was shown that $Mg(OH)_2$ was deposited at alkaline sites in the cathodically protected scratch[60,102].

Zn-based RP has also been used extensively on mild[116,117,121,122], carbon[77,123–126], and stainless steels [82,127–129] for galvanic protection in aqueous and marine atmospheric environments. ZnRP used on steels has shown the ability to perform as an effective sacrificial anode-based form of cathodic protection in marine environments[118,119,130–133]. Composite ZnRP pigment coating systems have been used before for the protection of carbon steel in marine environments[134–137]. These ZnRP have been shown to be an effective system for providing sacrificial anode-based cathodic protection to aircraft aluminum such as 2XXX and 5XXX-series aluminum alloys in marine environments[99,133,138,139]. The utilization of ZnRP on either steel or aluminum substrates results in the conversion of Zn to ZnO pigment via oxidation[58,133,140], which is considered "depleted" yet continues to provide barrier protection[126,134,141]. Zn is viewed as a "p-metal" and thought to have the character of a p-type semiconductor, whereas ZnO is a n-type semiconductor[142,143]. Thus, the combination of Zn–ZnO may form a p–n junction, which permits the flow of electrons and can control the electrochemical reaction of corrosion[143].

Another possible candidate for the replacement of chromated conversion coating is the use of aluminum-lithium passivation processes, including immersion in alkaline lithium salt solutions which has proven effective for aluminum and aluminum alloys[144]. Previous work has shown the formation of a stable corrosion-resistant film within a scribe defect that has stoichiometry of $Li_2[Al_2(OH)_5]_2 \cdot CO_3 \cdot nH_2O$ [145–147]. Lithium salts have been proposed as potential replacements for chromate-containing pigments in organic MRP coating systems[144]. Testing on AA 2024-T3 showed that the leaching lithium carbonates and lithium oxalates from an organic coating formed a protective layer in an artificial defect [145]. The effective corrosion protective properties of these layers were demonstrated by Visser et al. in a study based on electrochemical techniques[145–148]. However, none of this work has elucidated whether this substrate protection strategy can suppress IGC/IGSCC.

There has been limited research conducted on composite primers with multiple dissimilar pigments combined into a single coating. This shows another design parameter by which a coating can be tuned for the protection of the underlying substrate and suppression of IGC. The addition of multiple pigment chemistries within the same coating allows for a more robust protection scheme as the utilization (conversion of pigment), galvanic protection, and secondary barrier properties are a function of the pigment within the primer. The current literature on composite coatings is primarily focused on the combination of Mg, Zn, and their oxides mixed in various proportions to create composite primers [138,149]. The work of Shen et al. indicated enhanced sacrificial anode-based cathodic protection of a composite Mg + ZnO primer compared to MgRP analyzed on AZ91D magnesium alloy[150]. The secondary barrier properties of the composite Mg + ZnO primer were shown to form a robust passive layer with greater stability compared to MgRP[150]. Another study by Fayomi et al. investigated a Zn + MgO composite primer and found that the addition of MgO enhanced the corrosion performance of the coatings applied to mild steel[151,152]. There have also been efforts to understand the influence of additions of graphene and carbon nanotubes on both single pigment and composite coatings[132–135]. However, none of this work has elucidated whether this substrate protection strategy can suppress IGC/IGSCC. The limited work available in literature indicates a gap in the knowledge that merits further investigation into the electrochemical behavior of composite coatings compared to their single pigment primer counterparts.

This study focuses on a prospective alternative to chromate-based coating systems and single pigment primers by investigating a composite MgAlRP applied to a challenging substrate susceptible to IGC such as AA 7075-T651 free of any other inhibitors, pretreatment, and top coatings. The objective is to determine the relative performance of each system as a sacrificial anode, protection of the substrate, and examine the governing mechanisms. The performance of MgRP is used as a control in order to make an accurate determination as to the efficacy of the composite MgAlRP in providing sacrificial anode based cathodic protection to AA 7075-T651. The electrochemical behavior and coating performance of an Al-based MRP (AlRP) applied to AA 7075-T651 were reported in a previous publication[89]. Each MRP is evaluated using multiple electrochemical diagnostic techniques and characterization methods to identify performance and corrosion products formed. $E_{pit\,(\eta)}$ is used as a critical potential to determine susceptibility to EAC and IG-SCC. Moreover, these studies were combined with ASTM B117 exposure testing under salt fog. This initial work was conducted in NaCl but diagnostic experiments included variations in both $Cl^-$ concentrations and pH. Future work should consider wet/dry cyclic exposures, variation in pigment chemistry, and other relevant coating properties such as PVC.

## Experimental

### Materials: AA 7075-T651, MgRP, and MgAlRP

Peak aged and stress relieved AA 7075-T651 rectangular plates were machined to dimensions 1.5 mm thick, 200 mm long, and 76 mm wide via Mager™ high speed cut off saw. Samples were de-greased via alcohol bath and dried in lab air before the spray coating application. The composition of AA 7075-T651 is shown in **Table 1**. All bare uncoated samples were wet-polished to 1200 grit SiC paper until a mirror finish was obtained. The DEVCOM Army Research Lab (ARL) conducted spray-coating of the MgRP and composite MgAlRP according to the formulations shown in **Table 2** and applied to Milspec. The resin/pigment combination of the MgRP as produced by AkzoNobel (AN) Coatings (Amsterdam, Netherlands) and designation consists of epoxy resin and non-passivated 99.9% pure Mg pigment with flaked morphology at 26% PVC. The resin/pigment combination of the MgAlRP composite, shown in **Table 2**, is a combination of the AN MgRP and an Aluminum Rich Primer (AlRP) produced by Randolph Coatings (RC, Chicopee, Massachusetts US). The RC AlRP coating contains a spherical aluminum pigment alloyed with 5wt% Zn with spherical morphology. The combination of the AN MgRP with RC AlRP is possible due to the use of similar organic resin-hardener combinations with specifications listed in **Table 2**. The AA 7075-T651 panels area free of pretreatment. The MRP coated AA 7075-T651 samples do not include passivated pigments and do not include topcoats of any variety. Both intact and X-scribed defect MRP coated samples are tested. The MRP coated samples with X-scribed defect are made with the use of a tungsten carbide tip exposing the underlying substrate AA 7075-T651 in accordance with ASTM D1654 [73,74]. A bulk Al-5wt%Zn cylinder was produced, in effort to represent the Al-5wt%Zn pigment within the MgAlRP, via melting in a National Element electric resistance furnace conducted in ambient conditions. A graphite crucible was utilized, subject to 5 repeated melting cycles to achieve increased homogeneity. The $MgZn_2$ specimen was synthesized by the Kurt Lesker Company weighing between 3 to 6 grams. Conventional synthesis methods involve induction melting under vacuum with an argon-enrich atmosphere to mitigate volatilization and oxidation effects.

### Coating Characterization: Metal Pigment

Characterization of as-received, electrochemically tested, and accelerated environmental testing of MRP coating cross-sections were conducted with the use of scanning electron microscope (SEM) and energy dispersive X-ray spectroscopy (EDS) on a Quanta 650 system for imaging and elemental analysis. The SEM cross-sections are gathered under back scatter imaging (BSI) at a magnification of 500x, spot size of 5 nm, and an accelerating voltage of 10 kV at $10^{-8}$ torr to mitigate charging. Cross-sectioned MRP coating were mounted in epoxy and wet polishing in water to 1200 grit using an abrasive silicon carbide pad. The polished epoxy mounted MRP cross-sections were sputter coated with a conductive Au-Pd layer using a Cressington-108 at 30mA for 40 seconds with a standoff distance of 5 cm. The Au and Pd signal was excluded from EDS elemental analysis. The SEM BSI of each MRP cross-section was used to determine physical attributes (**Table 2**) such as average pigment size and coating thickness using 10 vertical thickness measurements on 3 pristine MgRP and MgAlRP cross-sections.

Metal rich primer crystalline composition and corrosion product identity were analyzed using X-ray diffraction (XRD). The Empyrean diffractometer XRD source used was a Cu-Ka source (1.54nm) with a 40-mA beam accelerated at 45 kV to perform continuous scans from 20–120° at a step size of 0.02. Previous work on Zn-rich primers with similar thickness demonstrated the presence of major Al peak at approximately 44.5º, which showed that the entirety of the MRP was being sampled [33]. Due to the pigment choice in the MgAlRP and the composition of AA 7075-T651 the Al-5wt.% Zn pigment peaks could not be differentiated from the substrate making the depletion of Al-rich pigment unattainable by XRD methods.

## Corrosion Electrochemistry Investigation on Bare Materials

The initial potentiodynamic polarization (PDP) scans of AA 7075-T651 (S-T rolling orientation), $MgZn_2$, MgRP coated AA 7075-T651, MgAlRP coated AA 7075-T651, 99.9% pure Mg, and an Al-5wt%Zn alloy were conducted in quiescent 0.6 M NaCl solution under full immersion conditions. The PDP provides the basis for assessing potential ranges in which IGC may be the predominant mechanism of SCC according to an anodic dissolution mechanism based on differences in critical potentials. The $E_{corr}$ evaluated by PDP and $E_{OCP}$ evaluated via OCP monitoring were recorded to examine galvanic relationships and potential relative to critical potentials. The anodic portion of the potentiodynamic scan started at 50mV below the OCP and scanned upward at a rate of 1mV/s. The cathodic leg of the potentiodynamic polarization scan started at 50 mV above the OCP and scanned downwards at a rate of 1mV/s. All electrochemical testing was conducted over a 0.785 $cm^2$ surface area.

The influence of [$Mg^{2+}$] concentration in the presence of $Cl^-$ on the electrochemical properties of AA 7075-T651 as well as Al-5wt%Zn was assessed by conducting PDP scans as well as 24-hour OCP monitors across a range of pH conditions from pH 3, unadjusted (UA, pH 5.8), pH 9, pH 10, and pH 11. The [$Mg^{2+}$] concentration was adjusted with the addition of $MgCl_2$ and titrated to the appropriate pH using stock solutions of 1 M HCl and 1 M NaOH. The influence of $Cl^-$ on the PDP behavior and 24-hour OCP at near neutral conditions was conducted separately using NaCl to avoid convoluting the influence of $Cl^-$ with the presence of $Mg^{2+}$ for bare AA 7075-T651, Al-5wt%Zn, $MgZn_2$, MgRP coated AA 7075-T651, and MgAlRP coated AA 7075-T651. The NaCl concentration is varied between 1, 10, 100, and 1000 mM at near neutral unadjusted pH conditions.

Additionally, PDP testing was conducted on AA 7075-T651, $MgZn_2$, Mg (99.9% pure), and Al-5wt%Zn in 0.6 M NaCl under quiescent full immersion conditions in unadjusted pH (5-5.5) as well as pH 11 to aid in understanding the galvanic coupling conditions of the intact and scribed coating scenarios in accordance with mixed potential theory. Galvanic couples of all types are mediated by the electron transfer kinetics between the anode, cathode, and the electrical and ionic resistances between the two. The pigments within MRP coating systems can be alloyed to achieve a certain charge capacity and balance between rapid sacrificial anode kinetics and maintain galvanically coupled potential ($E_{GC}$) below that of the substrate. The quantity of exposed sacrificial particles electrically connected to the substrate affects the coupled potential and the overall corrosion rate in the substrate. All electrochemical tests were conducted using a Gamry 600 potentiostat and a standard three-electrode cell containing a saturated calomel reference electrode (SCE) and a Pt counter electrode (CE); each specimen was used as the working electrode (WE).

**The UVA DC/AC/OCP Cycle Test: Charge Output and Barrier Assessment of Mg/MgAl-Rich Primers**

The MgRP and MgAlRP coating systems' performance as a sacrificial anode during cathodic prevention and electrochemical characteristics was evaluated under full immersion conditions in quiescent 0.6 M NaCl. The UVa DC/AC/OCP laboratory accelerated cycle test has been widely employed and holds significant value across various primer applications. This is due to its capability to adjust the PS hold potential to align with the desired galvanic couple potential and can be tailored to any specific alloy compositions and MRP choice [54,87,89,100]. The charge capacity of the primer can be assessed by integrating the current density vs time measured during the PS hold potential intended to represent the galvanic couple potential. The polarity of the current measured during the PS hold potential is an indication of the sacrificial anode cathodic protection a given MRP is able to offer. Extensive studies have been conducted on analogous MRP systems across a range of realistic exposure scenarios, encompassing 2xxx and 5xxx series aluminum alloys[54,58,87,89,90,94,98,100,102,107,113,153–156]. However, there is a lack reported data pertaining to 7xxx-series aluminum alloys.

The laboratory accelerated cycle testing method stands as a reliable means to assess a candidate coating's capacity for sacrificial cathodic protection, particularly in the presence of a macro-defect (scribe) exposing the underlying substrate [54,87,89,98,100,102]. This testing method is advantageous to alternative forms of exposure testing and yields greater mechanistic information that can be obtained instead of relying on pos-mortem characterization from accelerated environmental testing solely. The electrochemical characteristics pertaining to MRP coating performance in the form of charge capacity (and polarity), residual barrier impedance, and progression of the primer-substrate galvanic couple potential system after various states of discharge were recorded throughout DC/AC/OCP testing. The laboratory accelerated cycle testing involves a 14-cycle series of repeated OCP/EIS/PS hold procedure summing up to 100 hours of polarization time. The PS hold stage which imposes a galvanic couple potential for the MRP-substrate system was selected to be -0.95 $V_{SCE}$. This testing procedure includes a cumulative 15 hours of OCP collection time and 100 hours of PS polarization time. A detailed description of the test scheme and associated times is shown elsewhere[89]. Establishing a PS hold at -0.95 $V_{SCE}$ aims to evaluate each MRP's capability to discharge anodic current under conditions designed to guard against IGC (i.e. $E_{PS\ hold} < E_{pit\ (\eta)}$). All laboratory accelerated cycle testing was performed in triplicate to confirm reproducibility and ensure the trends shown are characteristic.

The extent of galvanic coupling between the MRP and AA 7074-T651 substrate was observed via the OCP stage of cycle testing as a greater pool of connected active MRP pigment is capable of suppressing coupled potential below the corrosion potential of AA 7075-T651. The OCP of the primer also gives a qualitative indication of the remaining pigment. The activation of metal rich primers occur at the most negative established OCP during cycle testing. According to mixed potential theory, the galvanic couple potential is influenced by the surface area of pigments, their polarization characteristics, and the exposed surface area between the bare AA 7075-T651 and the MRP coated AA 7075-T651[54,89,98,100]. In the following EIS stage of laboratory accelerated cycle testing the barrier properties of the intact primer and the remaining primer after discharge were examined. EIS testing is evaluated between $10^5$ and $10^{-2}$ Hz, 10 points per decade, and an AC amplitude of 65mV rms.

**Galvanic Corrosion Analysis using Mixed Potential Theory, PS Corrosion Electrochemistry, and Zero-Resistance Ammetry**

The evaluation of galvanic corrosion between the AA 7075-T651 substrate, mimicking a bare scratch, and the Mg/MgAlRP was performed using a zero-resistance ammeter (ZRA) test. Throughout the galvanic corrosion testing, an unexposed MgRP/MgAlRP coated AA 7075-T651 was galvanically coupled to a pristine bare AA 7075-T651 electrode. This galvanic coupling scenario simulates a scratch and forms a bimetal galvanic couple between the MRP coated AA 7075-T651 and the bare AA 7075-T651. This is advantageous as the galvanic coupled potential and galvanically coupled current density are formed naturally and not imposed or forced by a potentiostat. The galvanic couple potential is permitted to vary freely between the OCP of the primer and the exposed bare metal surface. The galvanic corrosion was monitoring over a 24-hour period of the dissimilar electrodes in quiescent 0.6 M NaCl. The bare 7075-T651 substrate with an exposure area of 0.785 $cm^2$ is connected as the WE, the MRP coated 7075-T651 operating as the CE, and an SCE is used as the reference electrode. In this setup, if a negative current on the bare 7075-T651 WE is measured this indicates electron flow entering the cathode. The indication that the MRP (CE) is acting as an anode is represented by a negative current measured over the WE. The distance between WE and CE was greater than 4 cm in the galvanic corrosion test which limits corrosion product transfer but does not limit electrochemical interactions. The galvanic coupling of the MRP coated AA 7075-T651 to bare AA 7075-T651 and ensuing electrochemical reactions may induce local pH change due to the dissolution of active pigment within the MRP coating. The local pH fluctuations were monitored with the use of a Mettler Toledo dual ISM pH microprobe positioned at a standoff distance of 5mm from each electrode surface. Galvanic couple testing was conducted in a 1:1 (MRP: bare 7075-T651) area ratio to assess the effect of a drop covering a scratch. Galvanic couple testing was also conducted in a 15:1 (MRP: bare 7075-T651) area ratio to simulate the conditions present near a scribe where a greater area of MRP is present than bare 7075-T651 substrate. All galvanic corrosion monitoring experiments were conducted in triplicate to ensure the trends shown are characteristic.

**ASTM B117 Q-Fog Salt Spray Testing**

DEVCOM ARL conducted accelerated life testing (ALT) on both intact and scribed MgRP and MgAlRP coated AA 7075-T651 panels, following ASTM B117 standards, using an Auto Technology salt fog chamber[157]. Testing occurred for a total duration of six weeks with three intact and three scribed MRP coated AA 7075-T651 panels removed from the camber at two-week intervals for characterization purposes. In order for comparisons to be drawn unexposed intact and scribed samples were set aside, as controls, to provide a baseline characterization profile. Once specimens were removed from the chamber at each two-week sampling period, they were promptly rinsed with deionized water and dried prior to handling and storage for characterization. Electrochemical testing that tracked the progression of OCP and EIS impedance behavior throughout the six-week exposure period to B117 ALT was conducted on intact MgRP and MgAlRP coated AA 7075-T651. Characterization was conducted in the form of SEM BSI cross-sections with EDS elemental mapping throughout the six-week exposure window taken to document the progression of damage profiles. The oxygen signal detected through EDS elemental mapping was used as a marker for the oxidation of pigment and substrate in both cross-section map scan and plan view with line scan profiles across the scribed region. The evaluation of scribe corrosion and protection against IGC involved washing ASTM B117 samples post six-week exposure in 50% nitric solution for two-three minutes. This was done to remove accumulated corrosion products from both the surface of the MRP and the interior of the scribe. Comparisons are drawn from control samples that were unexposed as well as six-week exposed panels without protection scheme to determine whether or not the candidate MRP coating were capable of preventing entirely or reducing both scribe corrosion and protecting against IGC.

## Results

### Characterization of As-Received Mg/MgAlRP coated AA 7075-T651

Characterization of as received MRP-AA7075-T651 cross-sections are conducted using BSI with EDS elemental mapping to assess primer thickness and particle size at 500x magnification. These physical characteristics of each coating can be in **Table 2** with cross-section micrographs shown in **Figure 1** for MgRP coated AA 7075-T651, and **Figure 2** for MgAlRP coated AA 7075. The MgRP coating contains flaked pigment of 26.5 $\mu m$ major axis and 12.2 $\mu m$ minor axis. The MgAlRP composite primer contains flaked Mg pigment of 18.2 $\mu m$ major axis and 6.6 $\mu m$ minor axis and spherical Al-5wt%Zn pigment with an average diameter of 9.8 $\mu m$. It should be noted that the MgRP and MgAlRP both contain the same AN MgRP pigment and resin combination and the difference physical dimensions of Mg pigment may be a result of sampling during characterization.

The micrographs in **Figure 1** and **Figure 2** show a well connected MRP coating layer that is capable of providing electrical connection throughout the thickness of the coating that is provided by the individual primer particles. However, not every particle is connected to the surrounding particles. The pigment volume concentration (PVC) has been studied elsewhere and is not the focus of this paper[92,116,158]. The total PVC of Mg pigment within the MgRP was 26% compared to the PVC of Mg pigment within the MgAlRP of 19%. This indicates a lower PVC for the Mg pigment in the composite MgAlRP than in the MgRP. However, if Al is activated, the MgAlRP has a greater combined PVC including Mg pigment (19%) and Al pigment (28%) with a gross Mg + Al pigment surface area of 14.5 $cm^2$ of the combined Mg and Al pigment per $cm^2$ of MgAlRP. This is contrasted with the gross surface area of 4.81 $cm^2$ for the Mg pigment per $cm^2$ of MgRP.

**Corrosion Electrochemistry Investigation**

Investigation of the polarization behaviors of bare AA 7075-T651, Al-5wt%Zn alloy, $\eta$ phase , MgRP coated AA 7075-T651, and MgAlRP coated AA 7075-T651 is shown in **Figure 3**. Recall the strengthening phase in 7xxx alloy is the $\eta$ phase which forms homogeneously in grain interiors and heterogeneously on the grain boundary. The PDP provides the basis for assessing potential ranges in which IGC occurs by dissolution of a boundary phase or zone. This is achieved by comparing critical potentials ($E_{corr}$, $E_{OCP,}$ $E_{pit}$) to ascertain potential ranges for IGC as discussed above. Such potential-dependent dissolution contrast may provide the predominant framework for IGC and IGSCC according to an anodic dissolution mechanism based on differences in critical potentials[39,53,54,89]. The AA 7075-T651 exhibits an $E_{corr}$ of -0.75 $V_{SCE}$ which appears to be very near its $E_{pit}$ (**Figure 3)**.

The η phase is seen to have an Ecorr of -1.0 $V_{SCE}$ with $E_{pit}$ at -0.86 $V_{SCE}$ (**Figure 3)**. The Al-5wt% Zn indicates an $E_{corr}$ of -0.94 $V_{SCE}$ which appears to be very near to its $E_{pit}$ (**Figure 3)**. The electrochemical theory of Galvele posits that IGC occurs when the applied potential, $E_{app}$, is such that $E_{pit\,(\eta)} < E_{app} < E_{pit,\,7075}$ or $E_{OCP,\,7075}$ [39,53]. One concept for cathodic prevention posits that protection is achieved when the grain boundary precipitate, $\eta$ phase, is cathodically protected below $E_{pit\,(\eta)}$[39,53]. Therefore, the potential range to expect IGC to occur is between -0.86 $V_{SCE} < E_{app} <$ -0.75 $V_{SCE}$.

The PDP of MgRP and MgAlRP coated AA 7075-T651 shifts the $E_{corr}$ to -1.14 $V_{SCE}$ and -1.06 $V_{SCE}$ respectively as shown in **Figure 3**. A 24-hour exposure at OCP is shown in **Figure 4** for bare AA 7075-T6511, $MgZn_2$, Al-5wt% Zn, and MgRP/MgAlRP coated AA 7075-T651. This shows the MgRP activates to a potential of -1.15 $V_{SCE}$ and maintains a potential of -1.07 $V_{SCE}$ up to the end of the 24-hour period (**Figure 4**). This potential shift may be sufficient to provide protection against IGC and IG-SCC as the potential is depressed below $E_{pit\,(\eta)}$ by the MgRP. This will depend on the galvanic couple potential attained. The MgAlRP is observed to attain a potential of -1.25 $V_{SCE}$ and remains below the OCP of bare AA 7075-T651 by the end of the 24-hour OCP monitor at -1.08 $V_{SCE}$. This potential shift is sufficient to provide protection against IGC and IG-SCC as the potential is depressed below $E_{pit\,(\eta)}$ by the composite Mg-Al-5wt% Zn pigment in the MgAlRP provided the galvanic couple potential remains near the OCP.

The effects of magnesium concentration, $[Mg^{2+}]$, and pH on the electrochemical properties of AA 7075-T651 are shown in **Figure 5**. The variation in the OCP of AA 7075-T651 is influenced by the $[Mg^{2+}]$, which decreases the $E_{OCP}$. However, there was a greater dependency noticed across varying pH (**Figure 5a**). Similarities are noticed with the $E_{corr}$ of AA 7075-T651 and can be seen to decrease to -1.3 $V_{SCE}$ in a basic pH 11 solution (**Figure 5b**). There is small variation (85mV) in the $E_{pit}$ of AA 7075-T651 as a function of pH (**Figure 5c**). However, all alkaline solutions are observed to have similar decreasing trend in $E_{pit}$ as a function of $[Mg^{2+}]$ (**Figure 5c**). The effect of magnesium concentration, $[Mg^{2+}]$, and pH on the electrochemical behavior of Al-5wt% Zn is shown in **Figure 6**. Both the variation in $E_{OCP}$ and $E_{corr}$ of Al-5wt% Zn can be seen to decrease as the solution becomes more alkaline (**Figure 6a and 6b**). The variation in $E_{pit}$ of Al-5wt% Zn decreases as $[Mg^{2+}]$ and pH increase (**Figure 6c**). The variation of the $E_{OCP}$ with $[Cl^-]$ independent from of the effects of $[Mg^{2+}]$ in near neutral solution can be seen for pristine bare AA 7075-T651, Al-5wt%Zn, $MgZn_2$, MgRP, and MgAlRP in **Figure 7**. This data implies that the relationship between key potentials suggested to be pertinent to IGC remain roughly similar across all $[Mg^{2+}]$ and pH levels tested.

The influence of $[Cl^-]$ on the electrochemical behavior of AA 7075-T651, Al-5wt%Zn, and $MgZn_2$ is shown in **Figure 8** at near neutral unadjusted pH conditions. The PDP of AA 7075-T651 in **Figure 8a** shows a decrease in the $E_{corr}$, and $E_{pit}$ with increasing $[Cl^-]$. The $E_{corr}$ can be seen to decrease approximately 300 $mV_{SCE}$ from a NaCl solution concentration of 1 mM to 1 M (in **Figure 8a)**. The Al-5wt%Zn shows similar trends to AA 7075-T651; however, with increasing $[Cl^-]$ the passive window vanishes and $E_{corr}$ decreases by 200 $mV_{SCE}$ between 1 mM and 1 M NaCl as seen in **Figure 8b**. The $MgZn_2$ possesses the lowest $E_{corr}$ at 1 M NaCl at -1.05 $V_{SCE}$ and decreases by 300 $mV_{SCE}$ from a NaCl solution concentration of 1 mM to 1 M NaCl as shown in **Figure 8c**. These results suggest that MgRP and MgAlRP have an OCP favorable towards sacrificial protection. Even at different $Cl^-$ concentrations, the relationships are preserved such that $E_{OCP, MRP} < E_{OCP,(\eta)} < E_{pit\,(\eta)} < E_{OCP,\ 7075}$.

The PDP behavior of AA 7075-T651 shows a to decrease in $E_{corr}$ from -0.76 $V_{SCE}$ at near neutral conditions (**Figure 9a**) to -1.3 $V_{SCE}$ under pH 11 conditions (**Figure 9b)** with the development of a passive region until $E_{pit}$ at -0.74 $V_{SCE}$ (**Figure 9b**). The bare Al-5wt%Zn alloy shows similar trends, with $E_{corr}$ decreasing from -0.97 $V_{SCE}$ at near neutral conditions (**Figure 9a**) to -1.28 $V_{SCE}$ with the development of a passive region under pH 11 conditions until pitting at -0.9 $V_{SCE}$ (**Figure 9b**). The $MgZn_2$ shows a decrease in $E_{corr}$ from -1.13 $V_{SCE}$ at near neutral conditions (**Figure 9a**) to -1.33 $V_{SCE}$ under pH 11 conditions until pitting at -0.85 $V_{SCE}$ (**Figure 9b**). The Mg displays a decrease in $E_{corr}$ from -1.55 at near neutral conditions (**Figure 9a**) to -1.65 under pH 11 conditions (**Figure 9b**). Furthermore, the mixed potentials describing the galvanic coupling of bare AA 7075-T651, Al-5wt%Zn alloy, $MgZn_2$, and pure Mg can be seen by the junctions formed in the PDP shown in **Figure**

**9.** The galvanic couple potentials of Mg and MgRP coupled to bare AA 7075-T651 were -1.47 $V_{SCE}$ and -0.8 $V_{SCE}$, respectively. The MgAlRP galvanic coupled potential was -0.75 $V_{SCE}$. The galvanically coupled potentials of MgRP and MgAlRP identified by PDP are static and do not represent how the galvanically coupled potential evolves over time. Judging from the E - log i data, polarization below $E_{pit\,(\eta)}$ requires a potential below -0.85 $V_{SCE}$ on AA 7075-T651 so that embedded $MgZn_2$ is polarized below $E_{pit\,(\eta)}$. Therefore, a long-term potential hold was conducted on bare AA 7075-T651 at potentials to assess the current density that must be supplied from the MRP to the AA 7075-T651 to attain a coupled potential sufficient to protect $E_{pit\,(\eta)}$.

The effect of 24-hour PS polarization in quiescent 0.6 M NaCl on the protection of AA 7075-T651 can be seen in **Figure 10a** for PS holds at of -0.95 $V_{SCE}$, -1.1 $V_{SCE}$, and -1.4 $V_{SCE}$. It is clear that the optimal potential is -0.95 $V_{SCE}$. This is done to determine whether the galvanic couples formed by each MRP:AA 7075-T651 galvanic couple is sufficient to provide protection to AA 7075-T651. It can be seen from **Figure 10a** that a PS hold of -0.95 $V_{SCE}$ and -1.1 $V_{SCE}$ requires 6 $\mu A/cm^2$ and 10 $\mu A/cm^2$ after 24 hours, respectively. The PS hold of -1.4 $V_{SCE}$ results in 2.6 $mA/cm^2$ after 24 hours, as seen in **Figure 10a**. Plan view optical characterization (**Figure 10b)** indicates negligible surface degradation is noticed for the -0.95 $V_{SCE}$ potential hold. The -1.1 $V_{SCE}$ potential hold results in minor surface degradation in the form of enhanced dissolution near stinger precipitates as seen in the plan view optical characterization shown in **Figure 10b**. The -1.4 $V_{SCE}$ potential hold resulted in major surface degradation in the form of gross-scale cathodic corrosion as seen in the plan view optical characterization shown in **Figure 10b**. These differences are best observed in the SEM BSI cross-section micrographs shown in **Figure 10c** where the cathodic corrosion progresses through the depth of the AA 7075-T651 sample at -1.4 $V_{SCE}$. It can be seen from **Figure 10a** that polarization to -0.95 $V_{SCE}$ and -1.1 $V_{SCE}$ requires a supply of 6 $\mu A/cm^2$ and 10 $\mu A/cm^2$ after 24 hours, respectively. The PS hold of -1.4 $V_{SCE}$ results in 2.6 $mA/cm^2$ after 24 hours. However, this latter potential is too severe and leads to high rates of $H_2$ evolution. The long-term anodic current density which must be supplied from the MRP to the AA 7075-T651 to attain a coupled potential of -0.95 $V_{SCE}$ is 6 $\mu A/cm^2$ (**Figure 10a**).

### The UVa DC/AC/OCP Cycle Test: Charge Output and Barrier Assessment of Mg/MgAl-Rich Primers

The cycle test was performed on intact pigmented coatings. The magnitude of the established OCP of intact MgRP throughout DC/AC/OCP cycle testing is shown compared to the 7075-T651 OCP (-0.75 $V_{SCE}$, dotted red line) in **Figure 11a**. The MgRP applied to AA 7075-T651 activated to an OCP of -1.53 $V_{SCE}$ on the first cycle (10 minutes of polarization). The MgRP activated to its most negative potential on cycle 1 and sustained an $E_{OCP}$ between $E_{app}$ until cycle 13 (75 hours) of the DC/AC/OCP cycle testing (**Figure 11a**). The MgRP at cycle 1 polarized to -0.95 $V_{SCE}$ or $E_{app} < E_{pit\,(\eta)}$ initially produced a net anodic current density of $+0.5\ \mu A/cm^2$ averaged over the total area of 0.785 cm$^2$ (**Figure 11b**). The peak current produced occurred in cycle 4 at $+\ 34\ \mu A/cm^2$ that decreased and remained at $+2\ \mu A/cm^2$ after 1000 seconds (**Figure 11b**). These are anodic current densities that contribute to protection of the AA 7075-T651, as the MgRP is the anode and the AA 7075-T651 substrate sits below its OCP. **Figure 11c** shows the Bode magnitude electrochemical impedance spectra collected throughout the DC/AC/OCP cycle testing. A cross-sectional EDS oxygen map can be seen in **Figure 11d** showing the oxidation profile of the post-DC/AC/OCP cycle testing MgRP sample. The oxygen EDS map shows the oxidation of the Mg pigment perimeter as well as the MgRP-AA 7075-T651 interface (**Figure 11d**).

The MgAlRP activated to an OCP of -1.15 $V_{SCE}$ on the first cycle (10 minutes of polarization) (**Figure 12a**). The MgAlRP activated to its most negative potential on cycle 4 and sustained an $E_{OCP}$ lower than -0.95 $V_{SCE}$ until cycle 11 (56 hours) of the DC/AC/OCP cycle testing (**Figure 12a**). The MgAlRP at cycle 1 polarized to -0.95 $V_{SCE}$ or $E_{app} < E_{pit\,(\eta)}$ initially produced a current density of $+7.5$ $\mu A/cm^2$ averaged over the total area of 0.785 cm$^2$ (**Figure 12b**). The peak current produced occurred in cycle 2 at $+\ 10.6\ \mu A/cm^2$ that decreased and remained at $+2\ \mu A/cm^2$ after 1000 seconds (**Figure 12b**). These are anodic current densities that do contribute to the protection of AA 7075-T651, as the MgAlRP is the anode. **Figure 12c** shows the Bode magnitude electrochemical impedance spectra collected throughout the DC/AC/OCP cycle testing. A cross-sectional EDS oxygen map can be seen in **Figure 12d** showing the oxidation profile of the post-DC/AC/OCP cycle testing MgAlRP sample. The oxygen EDS map shows the oxidation of both the Mg and Al-5wt% Zn pigment extending beyond the perimeter of the pigment with more uniform oxidation throughout the exposed pigment surface (**Figure 12d**). The electrochemical characteristics and performance of MgRP and MgAlRP throughout DC/AC/OCP cycle testing are shown in **Table 3**.

The remaining anodic charge capacity in these MRPs is not necessarily assessed because pigment particles may be electrically isolated by loss of particles in close proximity and passivation by oxide of hydroxide formation. It is suggested that OCP increase during cycle testing (**Figure 11a-12a**) occurs mostly due to passivation and electrically isolated pigment rather than substantial pigment loss, which is supported by the Backscatter Electron (BSE) micrographs taken on post-cycle testing MRP specimens (**Figure 11d, Figure 12d**). The MgRP shows partial Mg pigment oxidation because the interior of the Mg pigment is relatively unperturbed. Moreover, there is a layer of oxidation at the MRP-substrate interface (**Figure 11d**). In the MgAlRP, complete oxidation is apparent in Mg pigments as indicated by oxygen EDS signal through the entire thickness of the primer following the cycle test (**Figure 12d**).

The MgRP and MgAlRP barrier properties were assessed via PS EIS (at the primer OCP) intermittently throughout PS cycle testing to monitor impedance and coating defect area progression with increasing exposure time. The variation in low frequency (0.01 Hz) modulus of impedance can be seen for each cycle in **Figure 13a** with very little degradation in barrier properties limited to variation of 1 order or magnitude for MgRP over 100 hours of polarization at -0.95 $V_{SCE}$. The MgRP had $Z_{mod}^{0.01\,Hz}$ response of $1.9 \times 10^5\ ohm \cdot cm^2$ that decayed to $2.1 \times 10^4\ ohm \cdot cm^2$ after 100 hours of PS hold at -0.95 $V_{SCE}$. The MgAlRP is observed to have a higher $Z_{mod}^{0.01\,Hz}$ over 100 hours of polarization at -0.95 $V_{SCE}$ varying from $4 \times 10^6\ ohm \cdot cm^2$ to $1.3 \times 10^6\ ohm \cdot cm^2$ (**Figure 13a**).

The variation in OCP established at the end of each OCP cycle within DC/AC/OCP cycle testing can be seen in **Figure 13b**. The lowest potential obtained for MgRP throughout DC/AC/OCP cycle testing after cycle 3 was -1.25 $V_{SCE}$. The lowest potential obtained for MgAlRP throughout DC/AC/OCP cycle testing was in cycle 2 with a potential of -1.16 $V_{SCE}$ (**Figure 13b**). The MgRP produced the least charge as indicated in **Figure 13c** showing the charge accumulated throughout DC/AC/OCP for each cycle. A decrease in the charge density was observed after cycle 9 proceeding to a rise again after cycle 12 (**Figure 13c**). The MgAlRP produced the greatest charge density observed and unlike the MgRP, continued to increase in utilization over time (**Figure 13c**). Evaluation of primer galvanic couple kinetic behavior may provide further evidence towards understanding these differences in MRP performance.

The theoretical anodic charge capacity of each MRP, reported in **Table 3**, was assessed based on average primer thickness and volume per $cm^2$ (**Table 2**), the density of Mg / Al-5wt.%Zn, the molar volume, the exchange of two electrons necessary for Zn or Mg oxidation ($Zn^{2+}$ / $Mg^{2+}$) and/or the exchange of three electrons necessary for Al oxidation ($Al^{3+}$). For comparison, the maximum total anodic current output exhibited by each MRP during cycle testing was computed by integrating the anodic current output from each stage of PS potential hold at -0.95 $V_{SCE}$ (**Table 3**). The theoretical anodic charge (Q) output analysis demonstrated that MgRP has the lowest theoretical anodic Q capacity at 16.7 $C/cm^2$, while MgAlRP has a higher capacity at 27.2 $C/cm^2$ (**Table 3**). The maximum experimental anodic Q output for each specimen based on three series of cycle testing demonstrated that the MgAlRP has a maximum output of 0.5 $C/cm^2$ followed by the MgRP with 0.1 $C/cm^2$.

The fraction of experimental to theoretical anodic charge output demonstrated anodic charge usage (ACU) of 1.8% for MgAlRP, and 0.6% for MgRP, as shown in **Table 3**. Therefore, these primers have considerable protective capacity remaining following the cycle test, indicating that a considerable reservoir of Al-Zn and Mg remains available for local galvanic protection should a defect develop proximate to buried pigment. However, it should be mentioned that there is self-corrosion of MgRP and galvanic corrosion of MgAlRP consumes an unknown amount of charge rendering it unavailable. In summary, MgAlRP performs better than MgRP according to a number of metrics when polarized to -0.95 $V_{SCE}$ to mimic a galvanic couple with bare AA 7075-T651.

**Substrate – Primer Galvanically Coupled Potential and Current Densities**

Galvanic couples investigated consisted of intact MRP coating electrically connected to bare AA 7075-T651. The galvanic couple test is an excellent complement to the cycle test, as the galvanic potential spontaneously forms, is not static, and is not assigned. The coupled galvanic potential of MgRP and MgAlRP coupled to bare AA 7075-T651 in a 1:1 area ratio is shown in **Figure 14a** with the green dotted line denoting the OCP of AA 7075-T651 (-0.75 $V_{SCE}$). The coupled potential of MgRP and MgAlRP to AA 7075-T651 is -0.9 $V_{SCE}$; however, the MgAlRP is more stable and subject to fewer changes in the coupled potential as seen in **Figure 14a**. The galvanically coupled potentials shown in **Figure 14a** are free to evolve with time. The galvanically coupled current densities are shown in **Figure 14b** for MgRP and MgAlRP coupled to AA 7075-T651 in a 1:1 area ratio. The galvanic coupled current density of MgRP produces current spikes up to $+7.5\mu A/cm^2$ and decreases to below $1 \mu A/cm^2$ by the end of the 24-hour bi-metal galvanic couple. The MgAlRP produces current spikes up to $+7.5\mu A/cm^2$ and decreases to $+1.7\mu A/cm^2$ by the end of the 24-hour bi-metal galvanic couple. This is consistent with the current density of 6 $\mu A/cm^2$required to polarize bare AA 7075-T651 to -0.95 $V_{SCE}$.

The coupled galvanic potential of MgRP and MgAlRP to bare AA 7075-T651 in a 15:1 area ratio is shown in **Figure 15a** with the green dotted line denoting the OCP of AA 7075-T651 (-0.75 $V_{SCE}$). The coupled potential of MgRP and MgAlRP to AA 7075-T651 is -1.0 $V_{SCE}$; however, the MgAlRP is more stable and subject to fewer changes in the coupled potential, as seen in **Figure 15a**. Again, there are differences in the galvanically coupled potentials identified through PDP in **Figure 9a** as they are static points while the galvanically coupled potentials in **Figure 15a** are allowed to evolve over time. Comparing the galvanic coupled potentials of **Figure 14a** and **15a** there is a lower galvanic couple potential by 100 mV given larger MRP:substrate area ratios. The galvanically coupled current densities are shown in **Figure 15b** for MgRP and MgAlRP coupled to AA 7075-T651 in a 15:1 area ratio. The galvanic coupled current density of MgRP produces current spikes up to +10$\mu A/cm^2$ and decreases to 2.5 $\mu A/cm^2$ by the end of the 24-hour bi-metal galvanic couple. The MgAlRP produces current spikes up to +7.5$\mu A/cm^2$ and decreases to +5$\mu A/cm^2$ by the end of the 24-hourbi-metal galvanic couple (**Figure 15b**). This is consistent with the current density of 6 $\mu A/cm^2$ required to polarize bare AA 7075-T651 to -0.95 $V_{SCE}$.

The MRP CE alters the solution chemistry near the reacting electrode interfaces causing changes in the solution pH throughout the galvanic coupling experiments in both 1:1 and 15:1 area ratios as shown in **Figure 16**. The change in pH in front of the bare AA 7075-T651 WE can be seen in **Figure 16a**. Changes in pH were only observed when the area ratio 15:1 was over the WE. The MgRP exhibits an increase in the local pH near the bare AA 7075-T651 WE from an initial pH of 6 to 8.5-9.1 (**Figure 16a**). The local pH near the bare AA 7075-T651 WE changed from 6 to 8.0-8.5 in the case of MgAlRP. The MgRP CE in the 15:1 area ratio exhibits a slightly greater pH increase to a peak of 10.3 decreasing as low as 8.4 by the end of the 24-hour monitoring period (**Figure 16b**). In summary, it is seen that the pH shifts toward 10 at the bare AA 7075-T651 surface couple to MRP. This will affect activation and deactivation of Al-5wt%Zn, as changes in pH may correspond to regions in which dissolution of Al is favorable over solid oxide or hydroxide formation[159].

**Lab Accelerated Testing – ASTM B117 Salt Spray**

Intact and X-scribed MgRP and MgAlRP were tested in 0.6 M NaCl with a Q-Fog salt spray cabinet to assess scratch protection ability and performance of each MRP as a function of exposure time. Periodic OCP and EIS monitoring of the MRP coated AA 7075-T651 panels was conducted in 0.6 M NaCl tracking the progression of coating deterioration throughout B117 testing (**Figure 17**). The MgRP is shown to maintain protection potential at -1.1 $V_{SCE}$ over the course of six weeks of ALT (**Figure 17a** ). The composite MgAlRP was capable of maintaining an $E_{OCP}$ of -1.3 $V_{SCE}$ after six weeks of ASTM B117 salt spray testing in 0.6 M NaCl as seen in **Figure 17b**. It is also useful to compare the impedance behavior. The Bode magnitude electrochemical impedance spectra progression throughout the six weeks of accelerated laboratory testing can be seen for both MgRP and MgAlRP taken after 1-hour OCP monitoring in **Figure 17c** and **Figure 17d**, respectively. The MgRP at zero weeks in the as-received condition shows a $Z_{mod}^{0.01\,Hz}$ of $5.9 \cdot 10^5\ \Omega \cdot cm^2$ as seen in **Figure 17c**. The MgAlRP initially recorded a $Z_{mod}^{0.01\,Hz}$ of $2.8 \cdot 10^7\ \Omega \cdot cm^2$ as seen in **Figure 17d**. The MgAlRP sustains an order of magnitude greater $Z_{mod}^{0.01\,Hz}$ $(1.2 \cdot 10^6\ \Omega \cdot cm^2)$ compared to the MgRP with a $Z_{mod}^{0.01\,Hz}$ of $1 \cdot 10^5\ \Omega \cdot cm^2$ after six weeks of ASTM B117 testing. The phase angle progression can be seen for MgRP in **Figure 17e** and MgAlRP in **Figure 17f**.

Under paint and scratch corrosion behavior of the MgRP throughout ASTM B117 exposure are reported in **Figures 18** and **19** with BSI showed in conjunction with magnesium and oxygen signal from EDS. The oxygen signal is used to track corrosion damage herein. The intact MgRP is shown in **Figure 18** and can be observed to oxidize at the coating-electrolyte interface and at Mg pigment particles throughout the thickness of the intact MgRP coatings. The oxidation predominantly occurs on the pigments throughout the MgRP coating and not at the MRP-substrate interface (**Figure 18**). The scribed MgRP cross-section is shown in **Figure 19** and is seen to oxidize throughout the coating thickness as well as at the scribe wall (defect region). The scribed MgRP experiences an increase in magnesium and oxygen signal within the scribed region seen in **Figure 19** without MRP-substrate oxidation. The oxidized Mg pigments can be seen to reduce the Mg signal intensity (**Figure 19**).

Concerning mass transfer of pigment to the scratch, **Figure 20** shows BSI imaging and EDS line scans of oxygen and Mg for MgRP in both the 0-week as-received condition as well as after 6-weeks of ASTM B117 salt spray exposure testing. The as-received 0-week MgRP shows a high magnesium signal intensity within the intact region of the coating, with the alloying elements of AA 7075-T651 shown with increasing signal intensity within the scribed (uncoated) region (**Figure 20**). The post 6-week ASTM B117 condition reveals a reduction in the magnesium signal intensity within the intact region of the coating with an increase in the magnesium signal intensity within the scribed region by 3.5x. The magnesium signal intensity overlaps with the oxygen signal intensity implying the oxidation of magnesium within the MgRP. The EDS line scan is only capable of detecting AA 7075-T651 substrate within the scribed region to background levels ($< 200$ counts) suggesting the presence of magnesium and oxygen corrosion products. The 6-week ASTM B117 BSI micrographs shown in **Figure 20** show a significant coverage of magnesium and oxygen-rich corrosion products identified via EDS line scan throughout the scribed region, showing the ability to heal defected regions.

Under paint and scratch corrosion evolution of the MgAlRP throughout ASTM B117 exposure is reported in **Figures 21** and **22,** showed in conjunction with magnesium and oxygen signal from EDS. The oxygen signal is used to track corrosion damage herein. The intact MgAlRP is shown in **Figure 21** and can be observed to oxidize at the coating-electrolyte interface as well as throughout the thickness of the MgAlRP coating. However, the extent of oxidation in the MgAlRP is lower than that in the MgRP (**Figure 18)**. The oxidation is restricted to the Mg pigments throughout the MgAlRP coating, and does not appear in the Al-5wt%Zn pigments (**Figure 21**). There does not appear to be any MRP-substrate interface oxidation (**Figure 21**). The scribed MgAlRP is shown in **Figure 22** and is seen to oxidize throughout the coating thickness as well as at the scribe wall (defect region). The scribed MgAlRP has an increase magnesium and oxygen signal within the scribed region, as seen in **Figure 22**, without MRP-substrate oxidation. The increased Mg concentration is an indicator of transfer and redeposition of dissolved Mg released from the MgAlRP coating and transferred to the scribe. This indicates some ability of the MgAlRP to precipitate corrosion products on defected regions of the coating (scribe).

**Figure 23** shows BSI imaging and EDS line scans of MgAlRP in the 0-week as-received condition to serve as control when comparing oxidation effects to 6-week ASTM B117 salt spray exposure testing . The post 6-week ASTM B117 condition reveals a reduction in the magnesium signal intensity within the intact region and an increase in the magnesium signal intensity within the scribed region. The post 6-week ASTM B117 condition reveals a reduction in the aluminum signal intensity within the intact region and a decrease in the aluminum signal intensity within the scribed region. The overlapping magnesium and oxygen EDS signal shown in **Figure 23** are present throughout the intact coating as well as the scribed region. The 6-week ASTM B117 BSI micrographs shown in **Figure 23** exhibit good coverage of the corrosion products produced throughout the scribed region.

Corrosion products were tracked throughout ASTM B117 cycle testing on each MRP XRD analysis identified the composition and relative intensity of the crystalline phase seen in each primer (**Figure 24**). The MgRP can be seen to produce crystalline $Mg(OH)_2$ corrosion products as identified in **Figure 24a**, present in the 2- and 4- week MgRP samples with the 6-week $Mg(OH)_2$ peaks broadening into a weaker signal-to-noise ratio XRD spectra. The MgAlRP can be seen to produce crystalline $Al(OH)_3$, present in the initial condition, as well as a weak peak at 58° sharing position with the previously identified $Mg(OH)_2$. However, the peak at 52° fades into the background with continued ASTM B117 salt spray exposure (**Figure 24b**). It is worth noting that there may be additional corrosion products that are not detected within the diffraction spectrum either due to lack of crystallinity or due to sampling and detection limitations.

It is necessary to evaluate the ability of each primer to provide protection against scribe corrosion and IGC of the substrate which can be made by considering the cross-section BSI micrographs in **Figure 25**. In order to determine whether each coating can protect against scribe corrosion a control needs to be considered by which comparisons can be made. The pristine uncoated

AA 7075-T651 in both bare and scribed conditions serve as a control and can be seen in **Figure 25a** and **25b**, respectively. The six-week B117 exposure of AA 7075-T651 in the bare (**Figure 25c**) and scribed (**Figure 25d**) conditions without MRP protection scheme serve as a means of comparing the amount of degradation that occurs under equivalent conditions and exposure time. These two controls provide a means of comparing each MRP coating to a baseline as opposed to making qualitative judgements compared to one another. Comparing the pristine bare AA 7075-T651 (**Figure 25a**) to the six-week exposure of unprotected bare AA 7075-T651 to B117 salt spray testing (**Figure 25c**) shows dissolution of AA 7075-T651 up to a depth of 20 µm as well as what looks to be dissolution of IMC and fine hair-like cracks left behind characteristic of IGC. Comparing the pristine scribed AA 7075-T651 (**Figure 25b**) to six-week exposure of unprotected scribed AA 7075-T651 (**Figure 25d**) shows an increase in the scribe dimensions along the depth of the scribe by 10 µm and width of the scribe by 80 µm after six weeks of B117 testing. The cross-section BSI of MgRP coated AA 7075-T651 within the scribed region is shown in **Figure 25e** with no apparent dissolution through the thickness of the cross-section after six weeks of B117 testing. The cross-section BSI of MgAlRP coated AA 7075-T651 within the scribed region in **Figure 25f** does not show any signs of scribe deterioration. The deterioration of the bare AA 7075-T651 surface, widening of the scribe wall, and increase in the scribe depth can be attributed to increased dissolution due to the lack of MRP protection scheme. The indentation shown within the scribed region of the MgAlRP in **Figure 25f** originates from the scribe tool as the tip hardness is substantially greater than the aluminum substrate. These results may not fully represent the severity of corrosion as the field of view does not encompass a large portion of the sample in comparison to the area exposed and may be a conservative representation.

## Discussion

The present study indicates that the MgAlRP system is superior to the MgRP system. This is supported by suitably negative and stable OCP over time, dispenses high anodic charge, remains an anode in zero resistance ammeter testing, and possesses superior barrier properties measured via EIS. It is noted that pigment physical attributes are not equal as far as pigment volume concentration and surface area. The total PVC of Mg pigment within the MgRP was 26% compared to the PVC of Mg pigment within the MgAlRP of 19%. This indicates a lower PVC for the Mg pigment in the composite MgAlRP than in the MgRP. However, if the Al is activated, the MgAlRP has a greater combined PVC including Mg pigment (19%) and Al pigment (28%) with a gross Mg + Al pigment surface area of 14.5 $cm^2$ of the combined Mg and Al pigment per $cm^2$ of MgAlRP. The Mg pigment in MgAlRP possess different pigment dimensions (**Table 2**) with a surface area of 6.58 $cm^2$ of Mg pigment per $cm^2$ of MgAlRP. This is contrasted with the gross surface area of 4.81 $cm^2$ for the Mg pigment per $cm^2$ of MgRP. The combined Mg + Al pigment surface area of the MgAlRP is greater than the Mg pigment surface area in the MgRP by a factor of 3. The differences in physical attributes of the primer reflect a greater Mg surface area in the MgAlRP than in the MgRP by a factor of 1.37. This does not reflect considerations of electrically disconnected pigments or pigment fallout and assumes 100% utilization of pigment in each coating. As mentioned previously, the MgAlRP is a mixture of the AN MgRP and an AlRP produced by RC, therefore, the differences in Mg pigment dimensions may be a result of different sampling areas during scanning electron microscopy and not entirely reflective of the MgAlRP tested.

**Oxidation of Mg and Al in Hybrid MgAl-Rich Primer Systems**

The oxidation observed in the intact MgRP during cycle testing (**Figure 11d**) and ASTM B117 accelerated environmental testing (**Figure 18**) show similarities in the presence of partial oxidation of Mg pigment. This is apparent as the perimeter of flaked Mg pigment appears oxidized with an unoxidized interior. The oxidation of Mg pigment and formation of corrosion products can be considered to occur spontaneously in aqueous environments through the following electrochemical half-cell and overall reactions described in Equations 1-3 or 4:

$$2H_2O + 2e^- \rightarrow H_2 + 2OH^- \qquad 1$$

$$Mg \rightarrow Mg^{2+} + 2e^- \qquad 2$$

$$Mg + 2H_2O \rightarrow Mg(OH)_2 + H_2 \text{ (overall)} \qquad 3$$

$$Mg+2H_2O = MgO + H_2 \qquad 4$$

$$Mg(OH)_2 \leftrightarrow Mg^{2+} + 2OH^- \qquad 5$$

The chemical equilibrium between the $Mg^{2+}$ hydroxide and $Mg^{2+}$ and $OH^-$ in solution is described by Equation 5. The corrosion products identified in **Figure 24** are described by the above reactions and can be plotted to produce equilibria lines for the formation of stable Mg-based corrosion products in aqueous solution dependent on the initial amount of species present and the pH, as seen in **Figure 26**. This is described elsewhere[159]. The formation of solid $Mg(OH)_2$ requires a higher concentration of available $Mg^{2+}$ for the equilibrium formation of a stable corrosion product compared to the stability of solid MgO, as seen in **Figure 26**. The precipitated corrosion products present after

ASTM B117 are an indication of utilization of the pigment and of the alkaline conditions developed throughout the duration of testing.

The oxidation behaviors observed in the intact MgAlRP during cycle testing (**Figure 12d**) and ASTM B117 accelerated environmental testing (**Figure 21**) show similarities in the presence of complete oxidation of Mg pigment. The observed oxidation cross-section profiles are distinctly different than those observed in the MgRP coating (**Figure 11d** and **18**). It is a given that Al-Mg interaction may form a local galvanic cell between pigments as well as the substrate. Moreover, the dissolution of Mg within the composite MgAlRP coating changes the local pH under equilibrium conditions governed by reaction 5. The equilibrium pH describing the stability between $Mg^{2+}$ and $Mg(OH)_2$ for a $[Mg^{2+}]$ of $10^{-6}$ M is 11.3. This in turn impacts the relative corrosion product stability of $AlO_2^-$, stabilizing $AlO_2^-$ at pH 11-12[160]. Therefore, this could be a pathway for AlRP pigment activation unique to the MgAlRP system. It can be seen that at pH 11-12 that $AlO_2^-$ is more stable than either $Al(OH)_3$ or $Al_2O_3$ (**Figure 26**). This can be shown below.

Now, let us consider the oxidation and dissolution of Al described by the relevant electrochemical reactions for the formation of the corrosion products shown below by Equations 6-10. Starting with a neutral pH, Al is spontaneously oxidized in water to form $Al(OH)_3$ or $Al_2O_3 \cdot nH_2O$.

$$3H_2O + 3e^- \rightarrow 3H_{ads} + 3OH^- \quad (6)$$

$$Al + 3H_2O \rightarrow Al(OH)_3 + 3e^- + 3H^+ \quad (7)$$

$$2Al + 6H_2O \rightarrow 2Al(OH)_3 + 3H_2 \text{ (overall)} \quad (8)$$

$$Al^{3+} + 2H_2O \leftrightarrow AlO_2^- + 4H^+ \quad (9)$$

$$2Al + 3H_2O = Al_2O_3 + 3H_2 \quad (10)$$

The $Al^{3+}$ corrosion products identified in the above reactions can be plotted to produce equilibria lines for the formation of stable Al-based corrosion products in aqueous solution dependent on the initial amount of $Al^{3+}$ and $Mg^{2+}$ species present and the pH, as seen in **Figure 26** for $Al_2O_3$ and $Al(OH)_3$. The formation of $Al(OH)_3$ requires a lower concentration of available $Al^{3+}$ for the formation of a stable corrosion product than $Al_2O_3$, as seen in **Figure 26**. A mixed Al-Mg product has not been identified.

This process explains the origins of the cooperative or synergistic effect between the performance of Mg and Al. The local pH brought about by Mg oxidation in an Al/Zn/Mg system is speculated to affect the dissolution of Al. This is understood through the use of a chemical stability diagram for Al/Mg described by Santucci et al.[159]. Defects such as scribes may expose additional aluminum from the substrate where corrosion products are allowed to dissolve and form corrosion products as found in line EDS profiles showing an increase in the Mg signal within the scribed region in both scribed MgRP (**Figure 20**) and scribed MgAlRP (**Figure 23**). The corrosion product formation is hypothesized to play an important role in the protection of the AA 7075-T651 substrate during exposure to marine conditions; therefore, the conditions of stable product formation and dissolution trajectories will determine the corrosion product protection capacity of a particular MRP. To address this issue requires discussion of the effects of the pH on corrosion electrochemistry. The dissolution trajectory, or dissolution pathway toward equilibrium, of the composite primer is then altered by the participation of both pigments in the oxidation and corrosion process of MgAlRP, as shown in **Figure 26**. The dissolution trajectory of the Mg-Al can be seen in **Figure 26** and is shown to activate the Al as the dissolution of Mg increases the basicity of solution pulling the dissolution trajectory into a region of Al activity or stable $Al^{3+}$. The dissolution trajectory of Mg-Al crosses the equilibria line of $Al(OH)_3/Al^{3+}$ at a pH of 8.8 as opposed to the Mg dissolution trajectory remaining within the stable $Al(OH)_3$ region. This is the premise of enhanced electrochemical performance of the composite primer showing increased utilization of depassivated Al pigment.

**Electrochemical behavior of MRP – 7075-T651 galvanic coupling explaining the Mg-Al synergy**

Two findings must be discussed: (a) the increased utilization of the composite MgAlRP, and (b) the electrochemical differences of Al-5wt% Zn pigment within MgAlRP and bulk Al-5wt% Zn alloy. These can be explained with mixed potential theory and the corrosion thermodynamics of the governing electrochemical and chemical reactions written above. This analysis requires that pH be taken into consideration. Consider the dissolution of Mg within the MgRP below pH 11 for which reactions 1-5 are operative. Mg is unstable in water and spontaneously corrodes with water reduction resulting in $H_2$ evolution (reactions 1-2) and overall reaction 3. The pH rises due to the production of hydroxyl ions seen in reaction 1 which raises the local pH to an equilibrium pH of 10.4 established by the equilibrium pH of reaction 3.

The chemical effects of $[Mg^{2+}]$, $[Cl^-]$, and pH on the dissolution behavior of AA 7075-T651 and an Al-5wt% Zn alloy are shown in **Figure 5** through **Figure 8**. The Al-5wt% Zn alloy appears to show similar trends as AA 7075-T651 under the same pH range and $[Mg^{2+}]$ conditions; however, the Al-5wt%Zn alloy activates more readily at pH 10 than AA 7075-T651 (**Figure 6**). The AA 7075-T651 shows greater dependence of $E_{OCP}$ on pH for all [$Mg^{2+}$] with activation seen at pH 11 shown in **Figure 5**. The polarization behavior of AA 7075-T651, Al-5wt% Zn, and $MgZn_2$ (**Figure 8**) shows strong effects on lowering the $E_{corr}$ as a function of [$Cl^-$] independent of [$Mg^{2+}$] at near neutral conditions. It is important to examine whether AlRP can function as a sacrificial anode under these changes in pH, [$Mg^{2+}$], and [$Cl^-$].

The MgRP in the 1:1 area ratio initially responds by supplying a strong anodic current from the oxidation reaction of $Mg^0$ to $Mg^{2+}$ (reaction 2). The reaction is non-polarizable, remaining close to the Nernst potential associated with reaction 2. The cessation of dissolution activity decreases limiting the utilization of all Mg pigment within the MgRP **(Figure 14b).** The region separating the chemical reaction between $Mg^{2+}$ and $Mg(OH)_2$ is defined by the equilibrium of reaction 4. For an $Mg^{2+}$ concentration of $10^0 - 10^{-6}$ M the equilibrium pH for reaction 4 will vary between 8.4 and 11.6 [161]. This allows for the assessment of ion concentration within the aqueous electrolyte by measuring the pH near the reacting electrode surface.

An assessment of the dissolution of Mg via pH monitoring is shown in **Figure 16** for a galvanic coupling of the bare AA 7075-T651 and MgRP coated AA 7075-T651. The 1:1 and 15:1 area ratio reached a peak pH of 9.8 and 10.3 for the MgRP electrode interface corresponding to a $Mg^{2+}$ concentration in the electrolyte of $10^{-3}$ and $10^{-4}$ M, respectively as determined from the chemical stability modeling proposed by Santucci et al.[159] and shown in **Figure 26**. The 1:1 and 15:1 area ratio attained a pH of approximately 8.6 over the coating after 24-hour galvanic coupling to bare AA 7075-T651, substrate indicating an $Mg^{2+}$ concentration of $10^{-1}$ M at equilibrium with $Mg(OH)_2$. The sequence of reactions 1-5 describes spontaneous corrosion of the Mg pigment yielding aggressive self-corrosion of magnesium systems in the presence of water and NaCl. Similar results were reported by McMahon et al in the determination of Mg-Al synergy between the same MgRP and a MgAlRP[54].

Consideration should be made as to the fate of Al within the MgAlRP. Pure Al starting at pH 6 in NaCl solution is passive at $10^{-6}$ M $Al^{3+}$ with a minimum equilibrium solubility at pH 4.7, as shown by Santucci et al.[159]. The Al spontaneously passivates by half-cell reactions 1 and 2 to form $Al(OH)_3$ / $Al_2O_3$ and becomes polarizable. This can be seen in the work of McMahon et al., in which the $E_{GC}$ of an AlRP (Al-5wt%Zn pigment) polarizes the potential of a 5456 substrate (-0.8 $V_{SCE}$)[54]. The equilibrium chemical stability of $AlO_2^-$ with $Al(OH)_3$ / $Al_2O_3$ is governed by Equation 9. For a concentration of $10^{-6}$ M, $AlO_2^-$ is stable at pH 8 and above. McMahon et al. measured pH over an AlRP throughout the duration of a galvanic corrosion experiment in which an AlRP coated 5456 coupled to bare AA 5456 remained at a pH of 6.0-6.5, which is well within the thermodynamic stability region of $Al(OH)_3$ / $Al_2O_3$. It should be noted that $Al^{3+}$ cation buildup in solution has been noted to accelerate HER on Al alloys[162], which accelerates self-corrosion. The alloying of 5wt% Zn in the aluminum pigments cannot be ignored, as the stability of $Zn(OH)_2$ at pH 6 is also pertinent to the enhanced activity of MgAlRP. The formation of $Zn(OH)_2$ is thermodynamically stable at a 1 M $Zn^{2+}$ with a minimum solubility at pH 5.6[163].

Al is activated by the high pH, as suggested by the chemical stability modeling in **Figure 26**. The chemical stability modeling implies that Mg oxidation and the resulting pH rise to 8.6 thermodynamically activates the Al-5wt% Zn pigments (**Figure 26**). The production of hydroxyl ions as reaction by-products of Mg oxidation changes the electrolyte to pH 8.6 during the galvanic couple. The Mg-Al pigment oxidation was observed to shift the pH to 9 over the MRP in the 15:1 area ratio and 8.2 pH for the 1:1 area ratio which requires a very high concentration of $10^{-3}$ M for $Al^{3+}$ and $10^{-2}$ for $Mg^{2+}$ for $Al(OH)_3/Al_2O_3$ and $Mg(OH)_2$ to remain the stable species, respectively. This could explain how the Al-5%Zn pigment is activated to oxidize to $AlO_2^-$. This pH change renders the Al-Zn pigment susceptible to active dissolution as the dissolution trajectory is forced outside of the passive $Al(OH)_3$ stability region. Once this pH is achieved, Al is expected to oxidize to $AlO_2^-$ according to reaction 4, providing a second pathway to support the long-lasting cathodic protection achieved by the Al-5wt% Zn / Mg composite primer. These findings should be explored in a variety of other relevant environments and during wetting and drying typical of field exposures.

A simplified treatment of the galvanic coupling of MgRP and MgAlRP with AA 7075-T651 is illustrated in **Figure 27**. This treatment is not representative of all of the complexities within MRPs; however, serves the role of illustrating the influences of multiple pigments, resistive nature of the polymer matrix ($R_{polymer}$), and resistance at the surface of the MRP ($R_{surface}$). This treatment incorporates real polarization data collected in **Figure 9a** tested in unadjusted 0.6 M NaCl under quiescent condition. The limiting current density of AA 7075-T651 in 0.6 M NaCl under quiescent conditions is taken to be 1.2 x $10^{-5}$ A/$cm^2$ which was evaluated via finite element modeling (FEM) and experimental methods [164,165].The bi-metal galvanic couple formed between the Mg pigment in MgRP and the AA 7075-T651 is shown without resistance by the junction of the cathodic AA 7075-T651 in black and the Mg oxidation line in green (**Figure 27**).

The bi-metal galvanic couple formed between the Mg pigment in the MgAlRP and the AA 7075-T651 is shown by the junction of the cathodic AA 7075-T651 in black and the Mg oxidation line in blue (**Figure 27**). The Mg oxidation lines cross the cathodic AA 7075-T651 curve within the region dominated by the hydrogen evolution reaction (HER). The Mg oxidation line representing the bi-metal galvanic couple formed between MgAlRP and AA 7075-T651 experiences a larger quantity of current (charge) output due to the differences in Mg surface area between the Mg within MgRP (4.81 $cm^2$) and the Mg within MgAlRP (6.58 $cm^2$) which differs by a factor of 1.37. This illustrates that the increase in surface area of the Mg pigment does not have a significant impact on the galvanic couple formed between the Mg pigment in each coating and the AA 705-T651 substrate. The bi-metal galvanic couple formed between the Al-5wt%Zn pigment in the MgAlRP and the AA 7075-T651 is described by the junction of the cathodic AA 7075-T651 in black and the Al-5wt%Zn oxidation line in red (**Figure 27**). The oxidation of Al-5w%Zn crosses the cathodic portion of AA 7075-T651 within the current limiting oxygen reduction reaction (ORR) region. The galvanic couple formed between the Al-5wt%Zn and AA 7075-T651 does not have a great influence on the amount of charge supplied to the AA 7075-T651 substrate as the galvanic couple is limited by low mass transport-controlled ORR kinetics.

The net galvanic couple formed between the MgAlRP and AA 7075-T651 is shown in the dotted blue line as the sum of oxidation reactions occurring on both Mg pigment and Al-5wt%Zn pigment within the MgAlRP coating (**Figure 27**). This net galvanic couple for MgRP with AA 7075-T651 is represented by the yellow circle and similarly for MgAlRP with AA 7075-T651 by the pink circle, is relevant when there is no polymer present and ohmic contributions are negligible. As ohmic contributions become more relevant through the solution, polymer, and surface resistance this modifies the galvanic coupling formed at the interface between the AA 7075-T651 and the Mg pigment. The amount of resistance that is experienced between a given pigment within the MgRP or MgAlRP is dependent on the PVC and whether the pigment can be considered to be exposed to electrolyte or buried within the coating.

## Conclusion

Metal-rich primer-based cathodic protection was investigated to understand its viability in protecting AA 7075-T651 by achieving intermediate cathodic potentials to mitigate IGC and IG-SCC.

The guiding attribute considered to mitigate IG-SCC, MRP was $E_{couple} < E_{pit,MgZn_2}$ in neutral naturally aerated NaCl. The MgRP was capable of maintaining cathodic polarization of 100-150 mV below the OCP of AA 7075-T651 throughout galvanic coupling in 0.6 M NaCl. This potential meets the criteria of sacrificial cathodic prevention of 7075-T651 by maintaining potentials below $E_{OCP,7075}$ = −0.75 $V_{SCE}$ and $E_{pit,\eta}$ = − 0.85 $V_{SCE}$.

- Mg flake pigment in epoxy-based MRP rapidly activated and performed as a sacrificial anode. Mg piment was only partially expended as evident from scanning electron cross-sections. The cathodic protection criteria were met with the MgRP and achieved an activated potential of -1.5 $V_{SCE}$ on AA 7075-T651.
- The combination of Al-5wt% Zn pigment and Mg pigment in an epoxy-based MRP achieves intermediate cathodic potentials of approximately -1.1 $V_{SCE}$ on AA 7075-T651.
- The composite MgAlRP primer has shown enhanced galvanic protection, increased anodic charge output, and stable coupled potentials below the OCP of AA 7075-T651 suggesting that these MRPs may be utilized in static galvanic coupling condition on AA 7075-T651.
- The dissolution trajectory of the Mg-Al system is shown to have delayed activation the Al as the dissolution of Mg increases the basicity of the solution, pulling the dissolution trajectory into a region of Al activity or stable $Al^{3+}$. This is the premise of enhanced electrochemical performance of the composite MgAlRP system showing increased utilization.
- Composite MgAlRP systems are capable of (1) maintaining cathodic polarization of 200-250 mV below the OCP of AA 7075-T651 throughout galvanic coupling in 0.6 M NaCl, and (2) supplying positive current indicating the MRP is operating as intended with coating acting as anode.
- MRP utilization produces a pH increase associated with Mg oxidation that shifts the thermodynamic stability of $Al^{3+}$ to $AlO_2^-$. In this way, Al corrosion occurs spontaneously in NaCl solution.

## Acknowledgements

Research was sponsored by the Army Research Laboratory and was accomplished under Cooperative Agreement Number W911NF-20-2-20188 and the United States Airforce (USAF) under contract (#F7000-18-2-0006). The views and conclusions contained in this document are those of the authors and should not be interpreted as representing the official policies, either expressed or implied, of the Army Research Laboratory or the U.S. Government. The U.S. Government is authorized to reproduce and distribute reprints for Government purposes notwithstanding any copyright notation herein.

***Tables***

Table 1. Nominal composition of AA 7075-T651 and Al-5wt%Zn.

| AA7075-T651 | |
|---|---|
| **Element** | **Wt. %** |
| Al | Bal. |
| Zn | 5.1-6.1 |
| Cu | 1.2-2.0 |
| Mg | 2.1-2.9 |
| Fe | 0.5 |
| Si | 0.4 |
| Cr | 0.18-0.28 |
| Ti | 0.2 |

| Al-5wt%Zn | |
|---|---|
| **Elements** | **Wt. %** |
| **Al** | **Bal.** |
| **Zn** | **5** |

Table 2. Metal Rich primer systems included in this study.

| **Code** | **Primer Commercial Name** | **Provider** | **Resin** | **PVC (%)** | **Thickness (μm)** | **Pigment Diameter/ Dimensions (μm)** |
|---|---|---|---|---|---|---|
| MgAlRP (Epoxy,19/28%) | N/A | UVa-ARL | Epoxy | 28 (Al-Zn)<br><br>19 (Mg) | 46.2 ± 6.1 | 9.8 ± 4.8 (Al-Zn)<br>Length: 18.2 ± 6.2 (Mg)<br>Width: 6.6 (Mg) |
| MgRP (Epoxy,26%) | Aerodur 2100 | AN | Epoxy | 26 | 44.5 ± 6.4 | Length: 26.3 ± 8.4<br><br>Width: 12.2 ± 4.4 |

AN = AkzoNobel, PVC = pigment volume concentration.

Table 3. Metal rich primer charge capacities and cathodic protection performance

| **Metal-Rich Primer** | **Theoretical Anodic Q from MRP ($C/cm^2$)** | **Maximum Experimental Anodic Q Output ($C/cm^2$) by End of Cycle Test** | **Anodic Q Usage by End of Cycle Test** | **Initial Activated OCP** | **OCP Upon Completion of the Cycle Test** | **Average Scribe Width** (µm) |
|---|---|---|---|---|---|---|
| MgAlRP (Epoxy, 19/28%) | 27.2 | 0.5 | 1.8% | -1.15 ± 0.06 $V_{SCE}$ | -0.9 ± 0.02 $V_{SCE}$ | 171 ± 15 |
| MgRP (Epoxy,26%) | 16.7 | 0.1 | 0.6% | -1.5 ± 0.04 $V_{SCE}$ | -0.95 ± 0.03 $V_{SCE}$ | 225 ± 20 |

Q = charge, C = Coulombs, XRD = X-Ray Diffraction, OCP = open circuit potential, Activated OCP = stabilized OCP following immersion and sufficient coating wetting, MRP = metal-rich primer

## Figures

## Figure 1

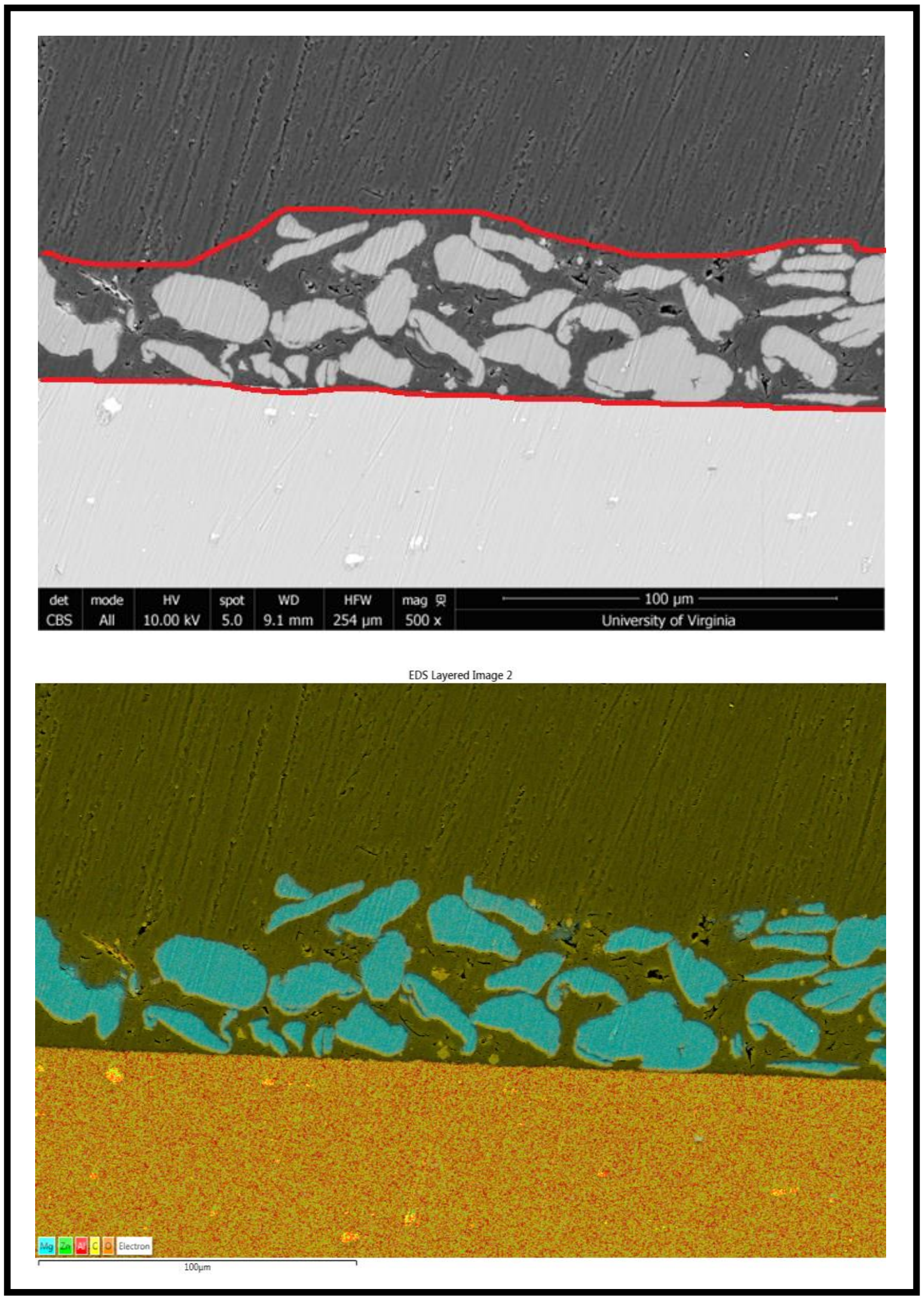

Figure 1. **a)** Cross sectional BSI electron micrographs and **b)** EDS map scan of pristine unexposed MgRP applied on AA7075-T651.

**Figure 2**

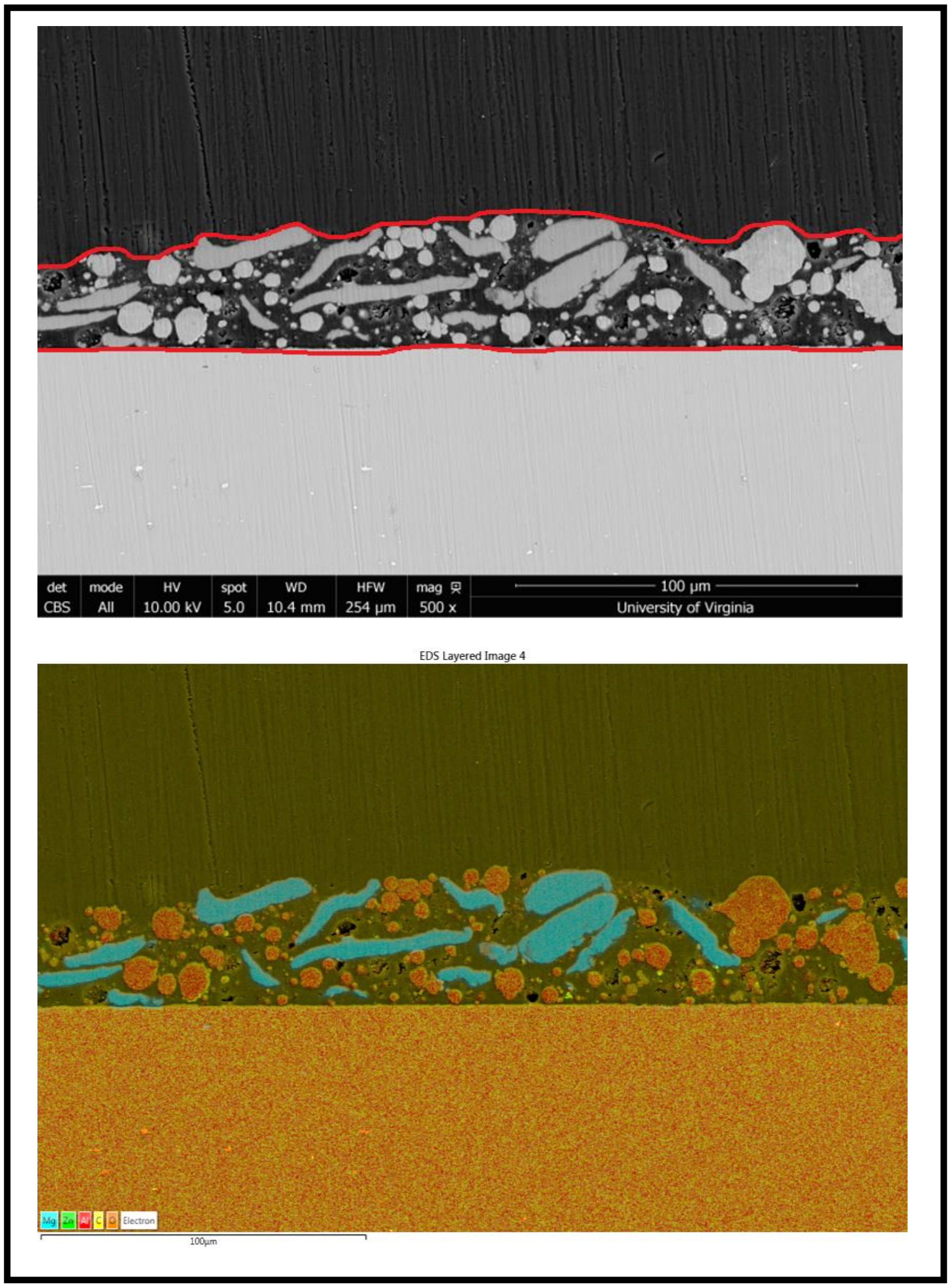

Figure 2. **a)** Cross sectional BSI electron micrographs and **b)** EDS map scan of pristine unexposed MgAlRP applied on AA7075-T651.

**Figure 3**

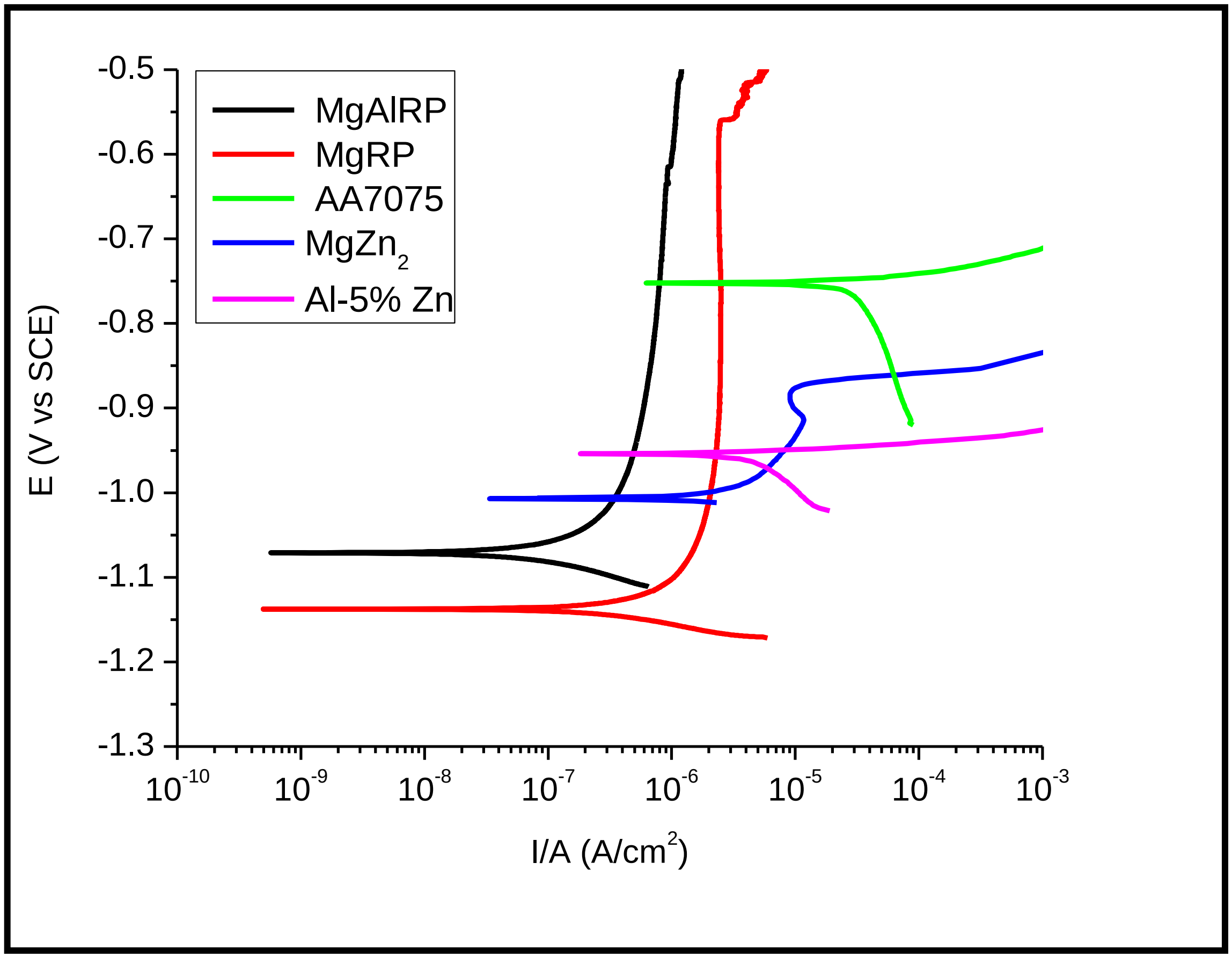

Figure 3. Potentiodynamic polarization scan of bare AA7075-T651, $MgZn_2$, Al-5wt%Zn, MgRP, and MgAlRP coated AA 7075-T651 under full immersion conditions in quiescent 0.6 M NaCl.

**Figure 4**

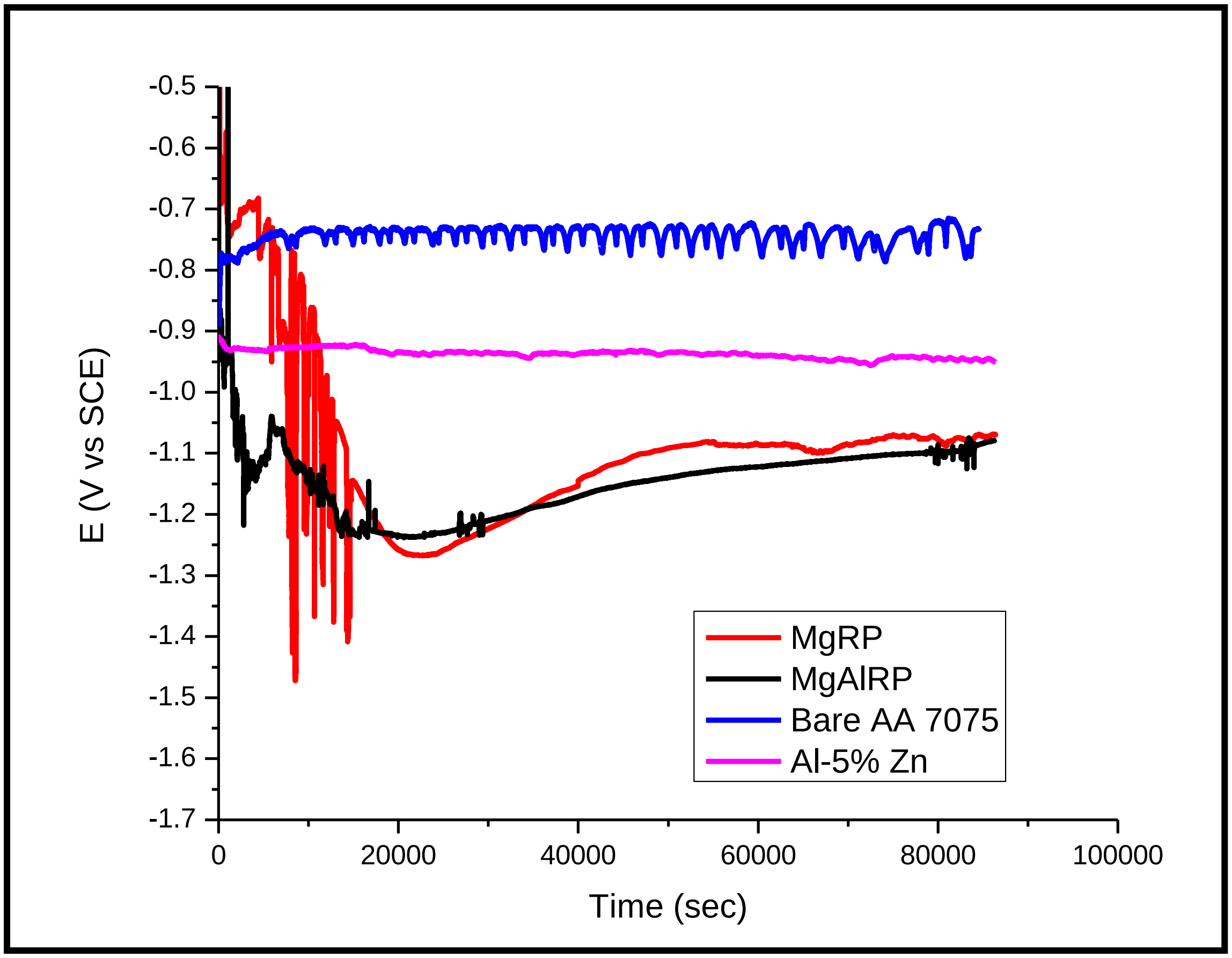

Figure 4. Long term open circuit potential of bare AA 7075-T651, Al-5wt% Zn, MgRP, and MgAlRP coated AA7075-T651 under full immersion conditions in quiescent 0.6 M NaCl.

**Figure 5**

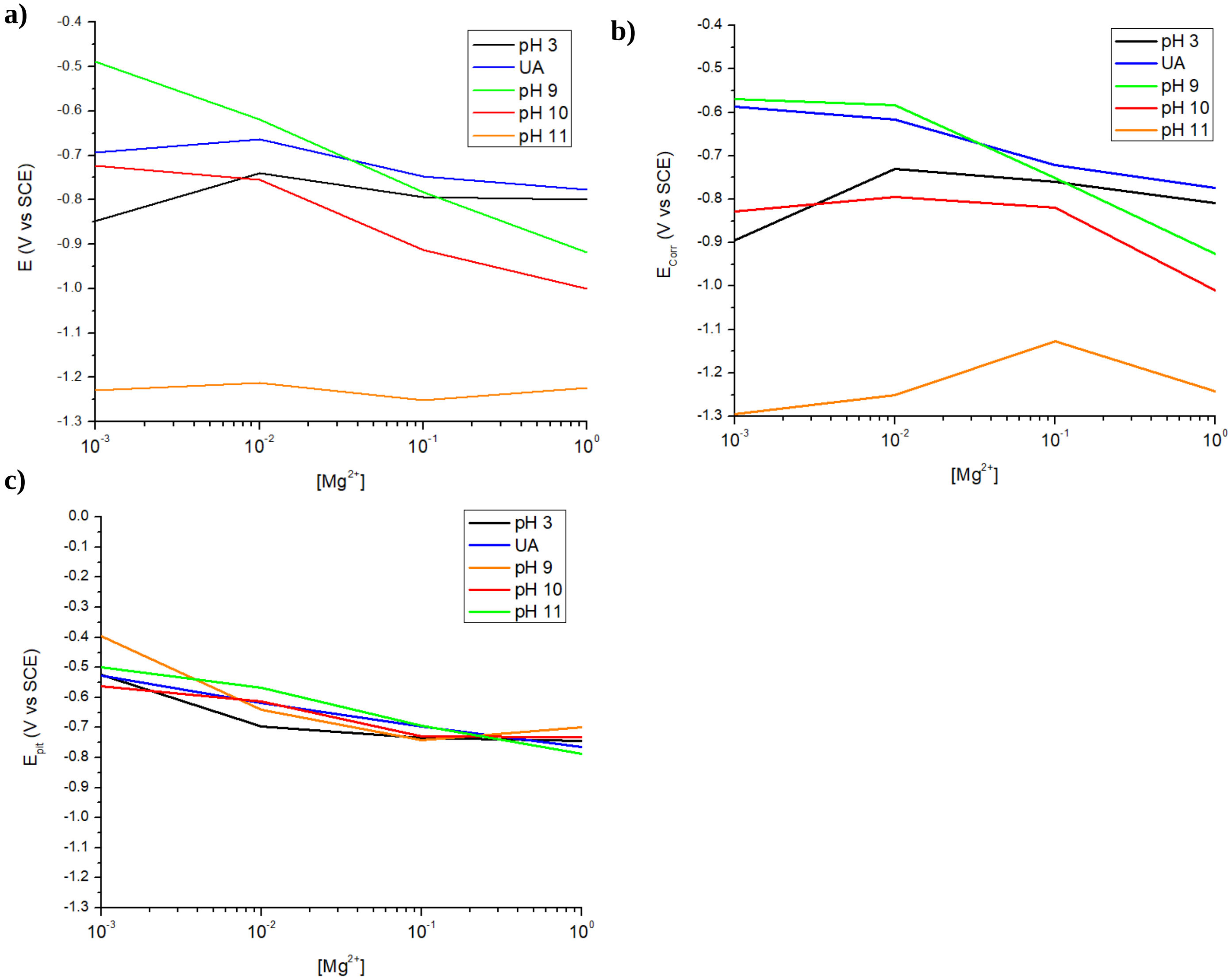

Figure 5. Effect of $[Mg^{2+}]$ in the presence of $Cl^-$ and influence of pH on the electrochemical behavior of AA 7075-T651 with **a)** the post 24-hour open circuit potential, **b)**the corrosion potential obtained through potentiodynamic polarization, and **c)** the pitting potential obtained through potentiodynamic polarization. Electrochemical testing was conducted under quiescent full immersion conditions. Variation in $[Mg^{2+}]$ is achieved with $MgCl_2$ and pH is adjusted via titrations of NaOH and HCl for either basic or acidic conditions.

**Figure 6**

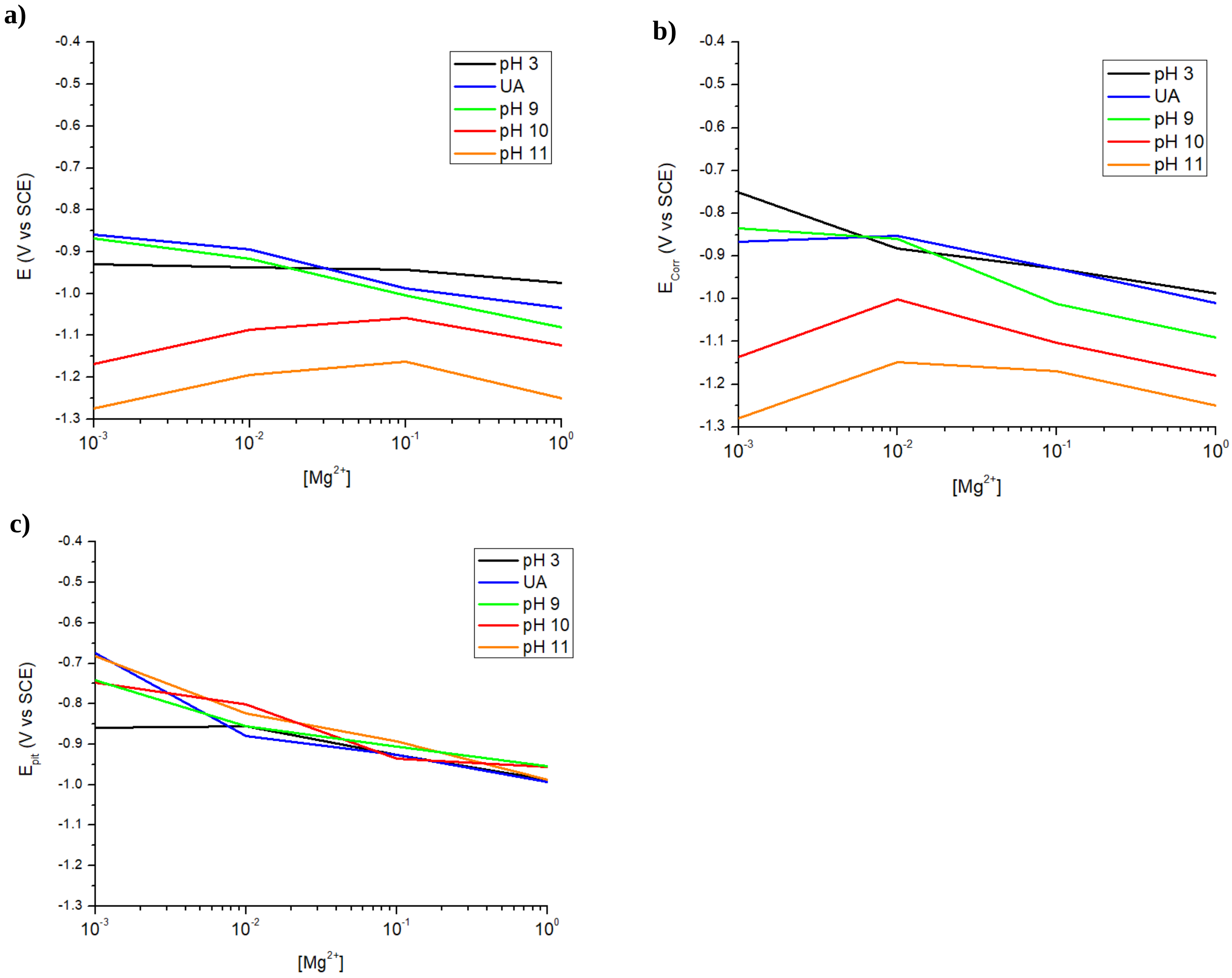

Figure 6. Effect of [$Mg^{2+}$] and influence of pH on the electrochemical behavior of **a)** Al-5wt%Zn bulk alloy with the post 24-hour open circuit potential, **b)** the corrosion potential obtained through potentiodynamic polarization, and **c)** the pitting potential obtained through potentiodynamic polarization. Electrochemical testing was conducted under quiescent full immersion conditions. Variation in [$Mg^{2+}$] is achieved with $MgCl_2$ and pH adjusted via titrations of NaOH and HCl for either basic or acidic conditions.

**Figure 7**

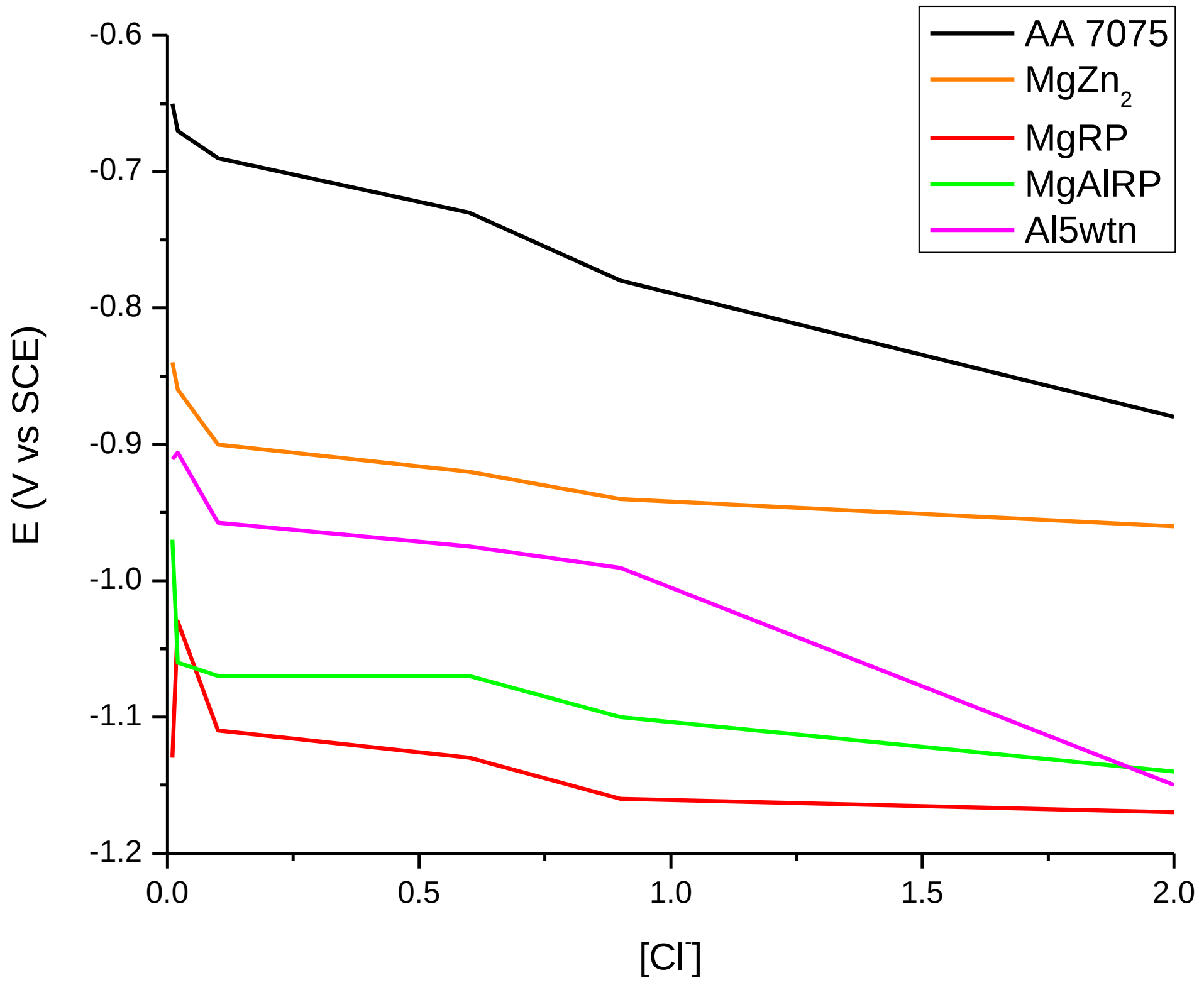

Figure 7. Effect of [$Cl^-$] on the post 24-hour open circuit potential of bare AA 7075-T651, $MgZn_2$, Al-5wt%Zn alloy, MgRP, and MgAlRP. Electrochemical testing is conducted under quiescent full immersion conditions. Variation in [$Cl^-$] is achieved with NaCl at near neutral conditions.

**Figure 8**

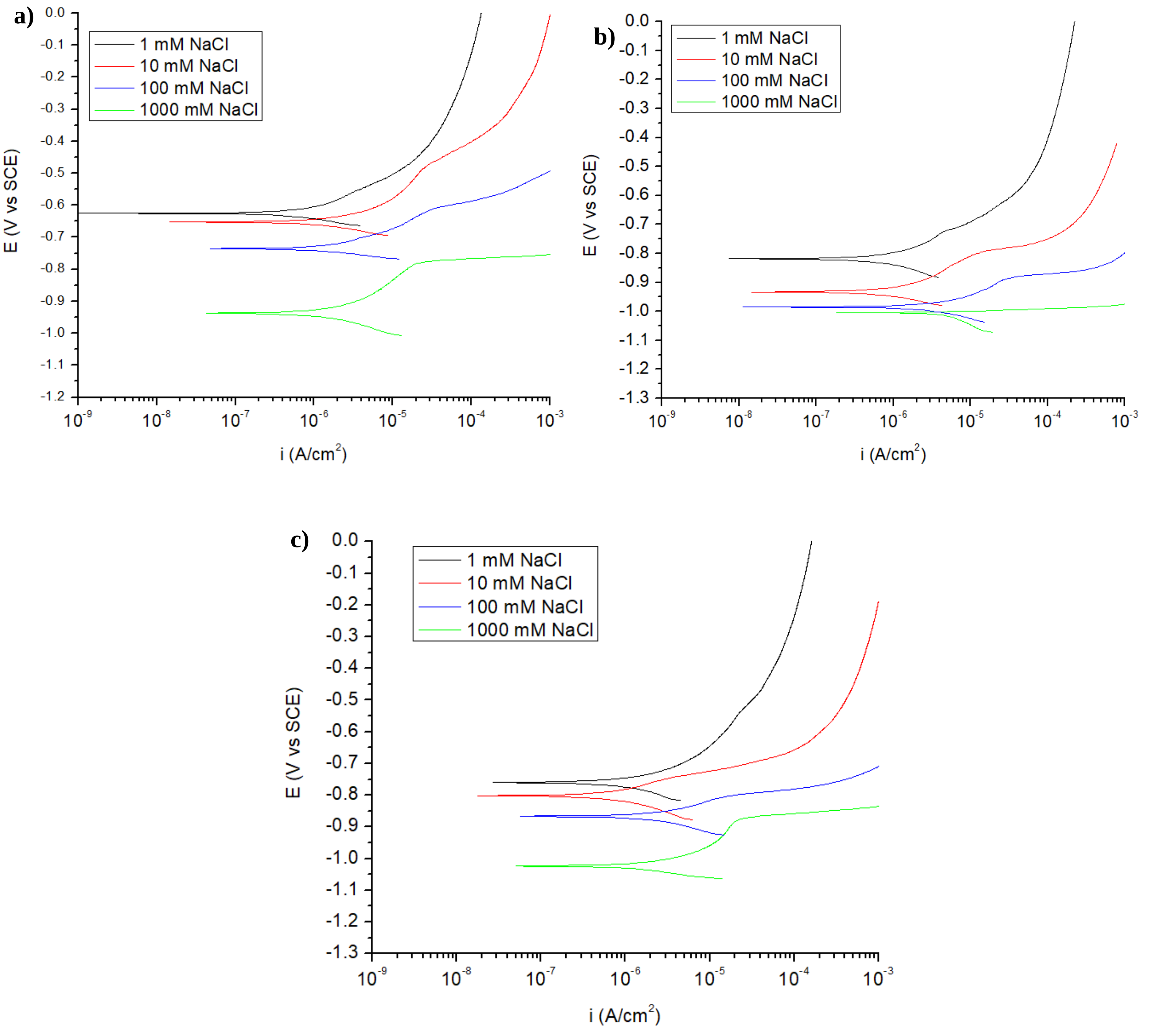

Figure 8. Influence of [$Cl^-$] on the electrochemical behavior of **a)** AA 7075-T651, **b)** Al-5wt% Zn alloy, and **c)** $MgZn_2$. Electrochemical testing is conducted under quiescent full immersion conditions. Variation in [$Cl^-$] is achieved with NaCl at near neutral conditions.

**Figure 9**

**a)**

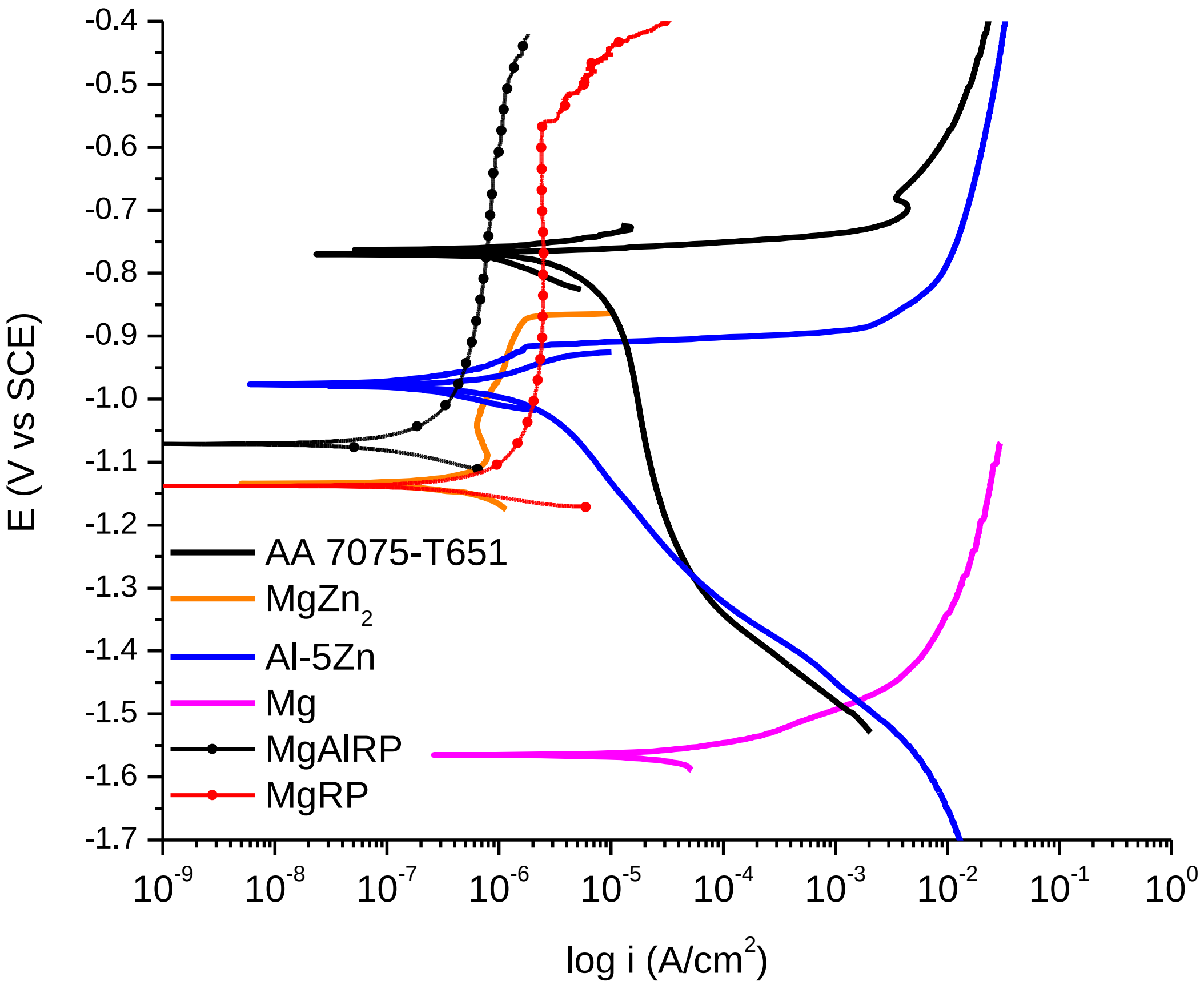

**b)**

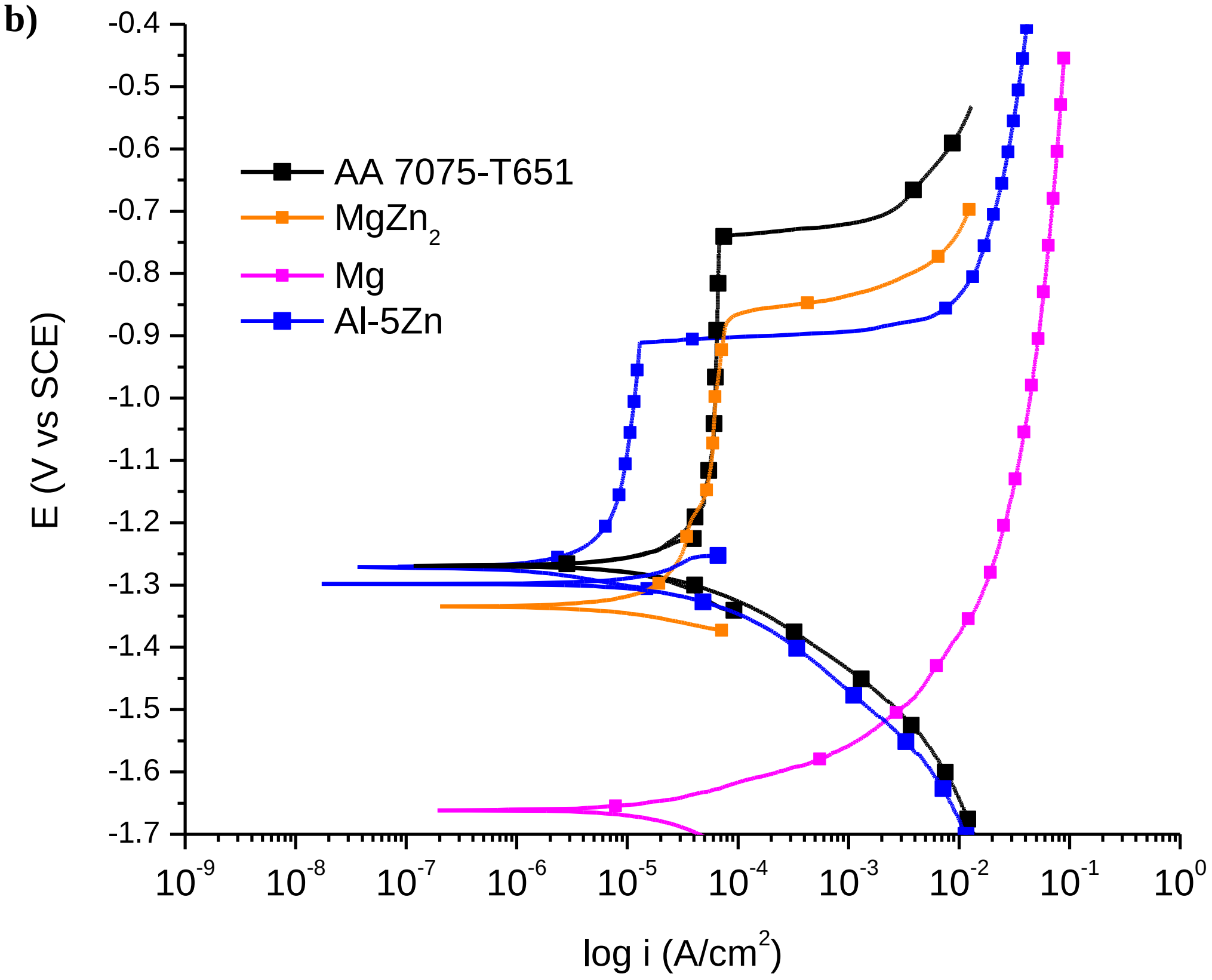

Figure 9. Potentiodynamic polarization diagram for AA 7075-T651, $MgZn_2$, Al-5wt%Zn alloy, and Mg (99.9%) in quiescent 0.6 M NaCl under full immersion in **a)** unadjusted pH (5-5.5), and **b)** pH 11. Solution pH adjustments are conducted with NaOH.

**Figure 10**

a)

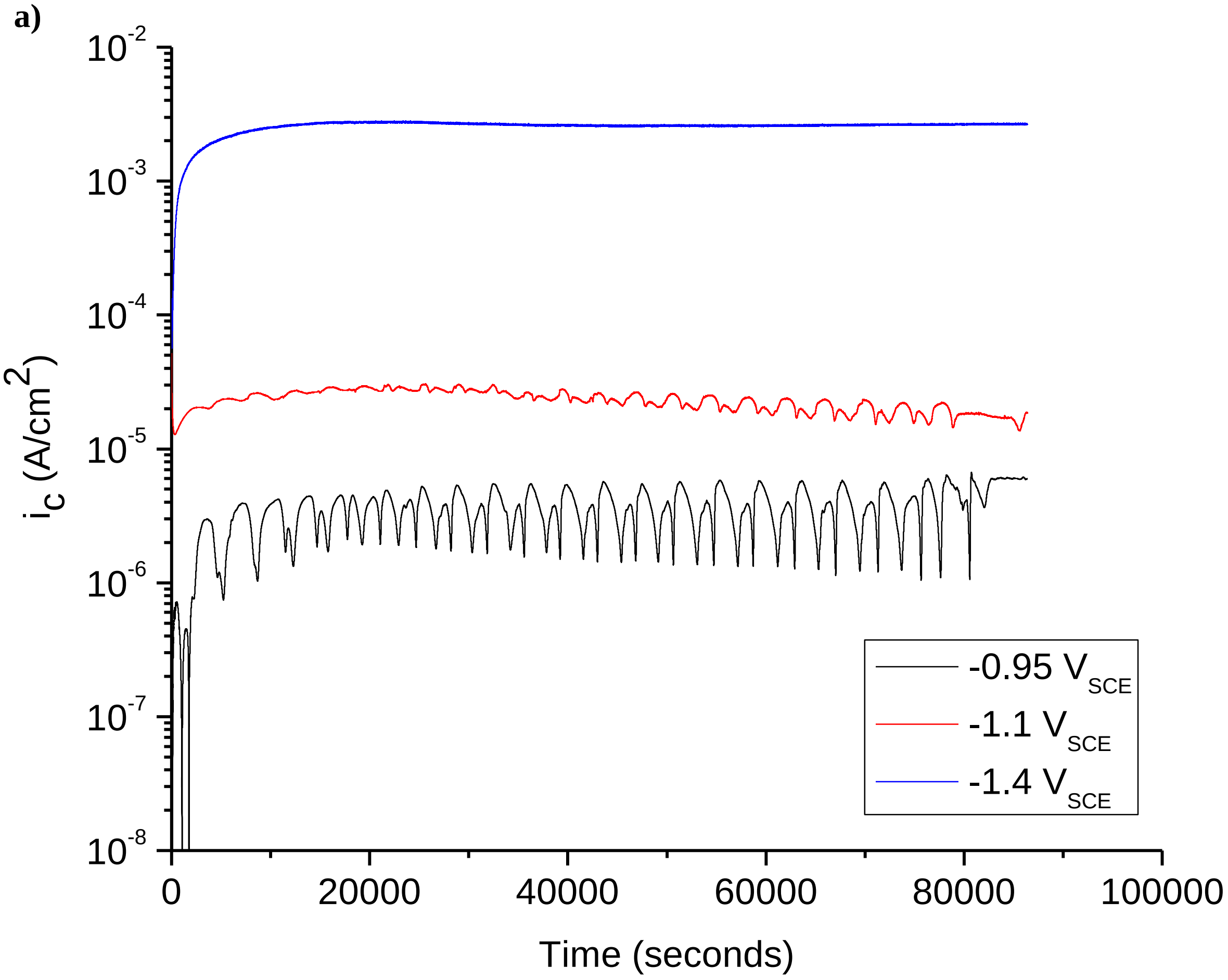

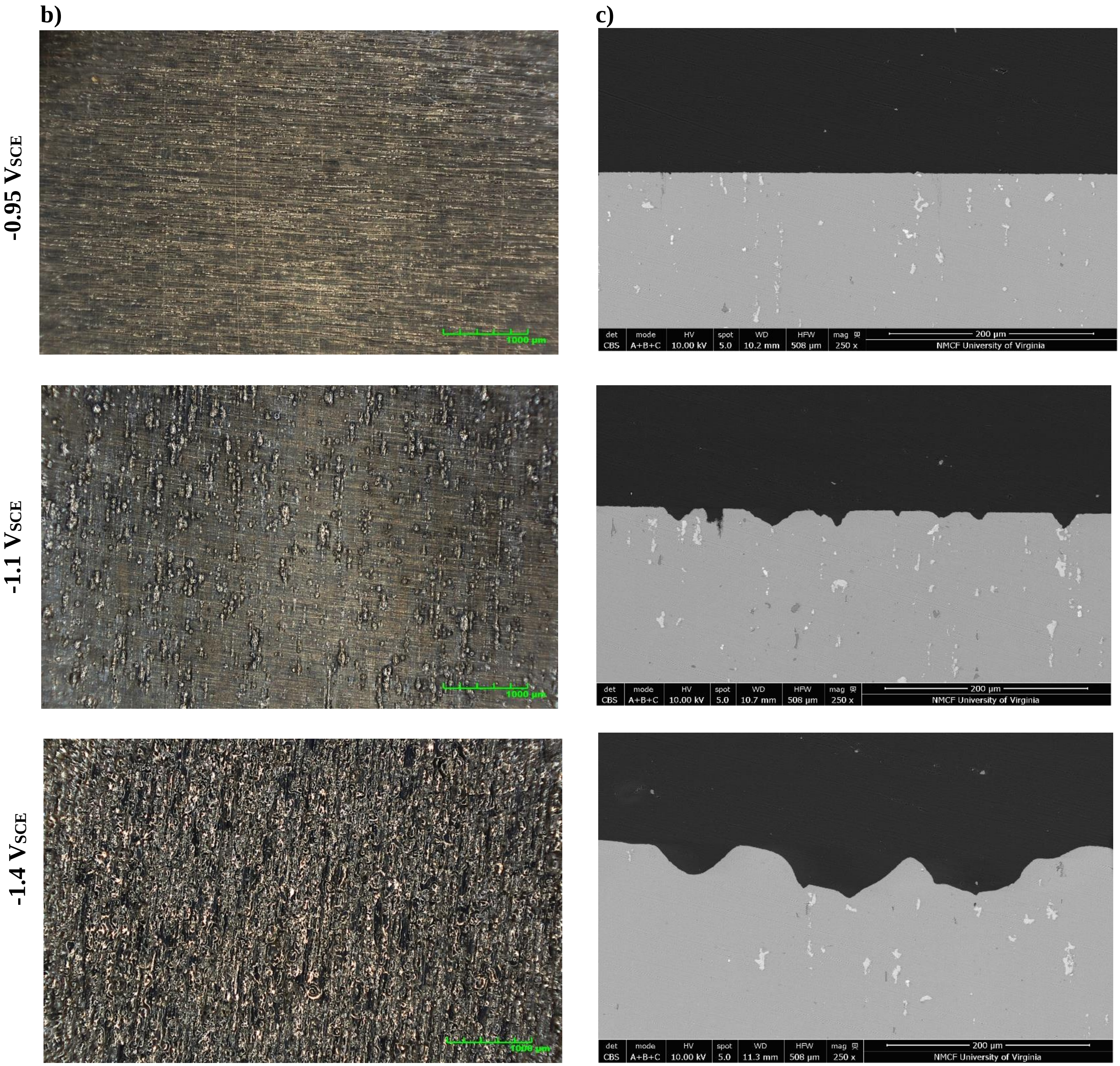

Figure 10. **a)** Potentiostatic hold on pristine bare AA7075-T651 at DC potential holds of -0.95$V_{SCE}$, -1.1 $V_{SCE}$, and -1.4 $V_{SCE}$ for 24-hours. Testing is conducted under full immersion conditions in quiescent 0.6 M NaCl. **b)** Plan view optical microscopy for each DC potential hold. **c)** SEM BSI cross-sections showing penetration through the depth of the sample as a function of DC polarization.

**Figure 11**

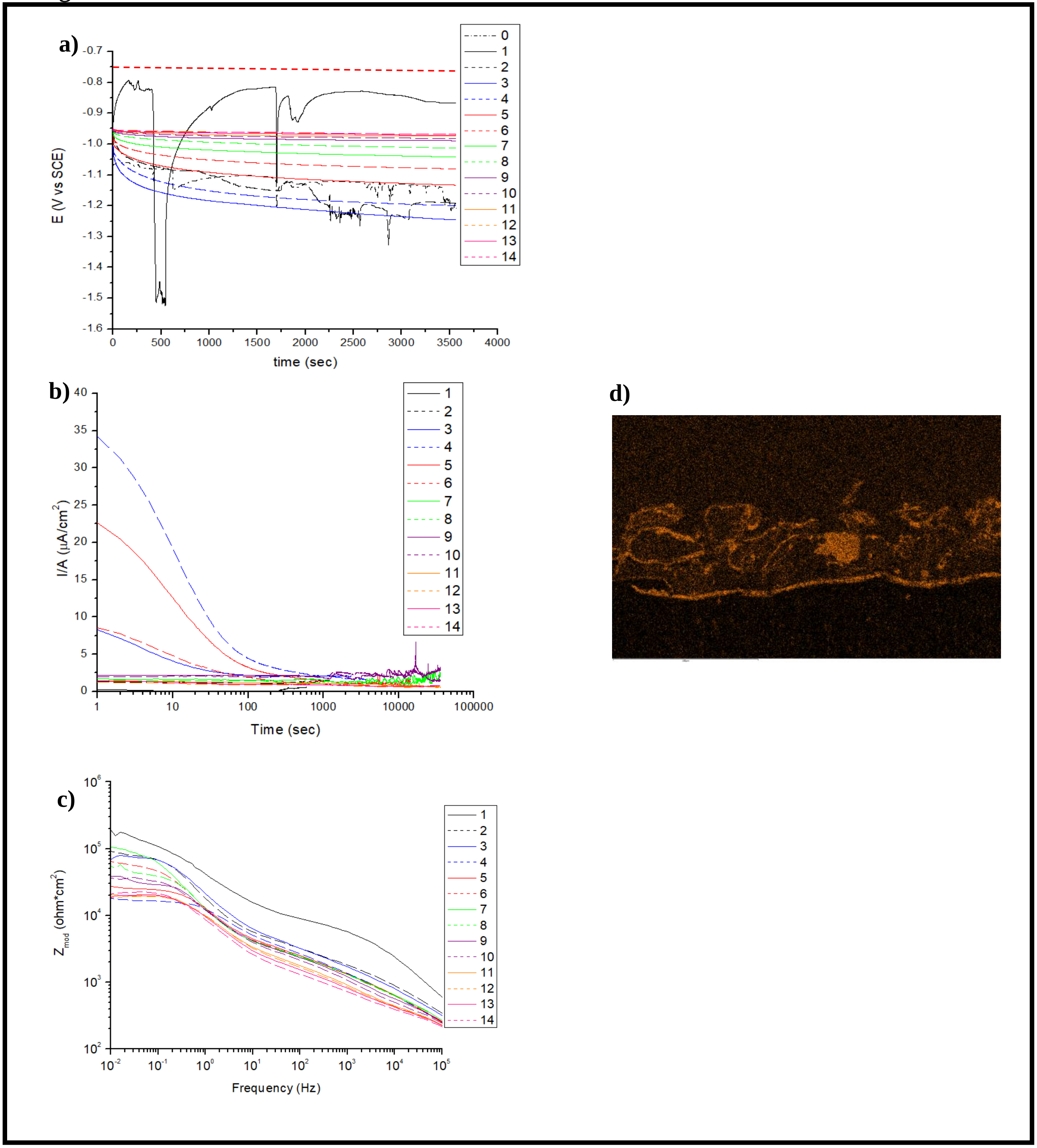

Figure 11. OCP/AC/DC cycle testing of MgRP applied to AA7075-T651 with DC potential hold at -0.95 $V_{SCE}$ and the legend denoting the cycle. **a)** Open circuit potential is shown with a red dotted line denoting the OCP of AA7075-T651, **b)** the current density output, **c)** with residual

barrier properties shown in the Bode impedance response, and **d)** the oxygen EDS cross-section of MgRP after 100 net hours of DC potential hold at -0.95$V_{SCE}$. Testing is conducted under full immersion conditions in quiescent 0.6 M NaCl.

**Figure 12**

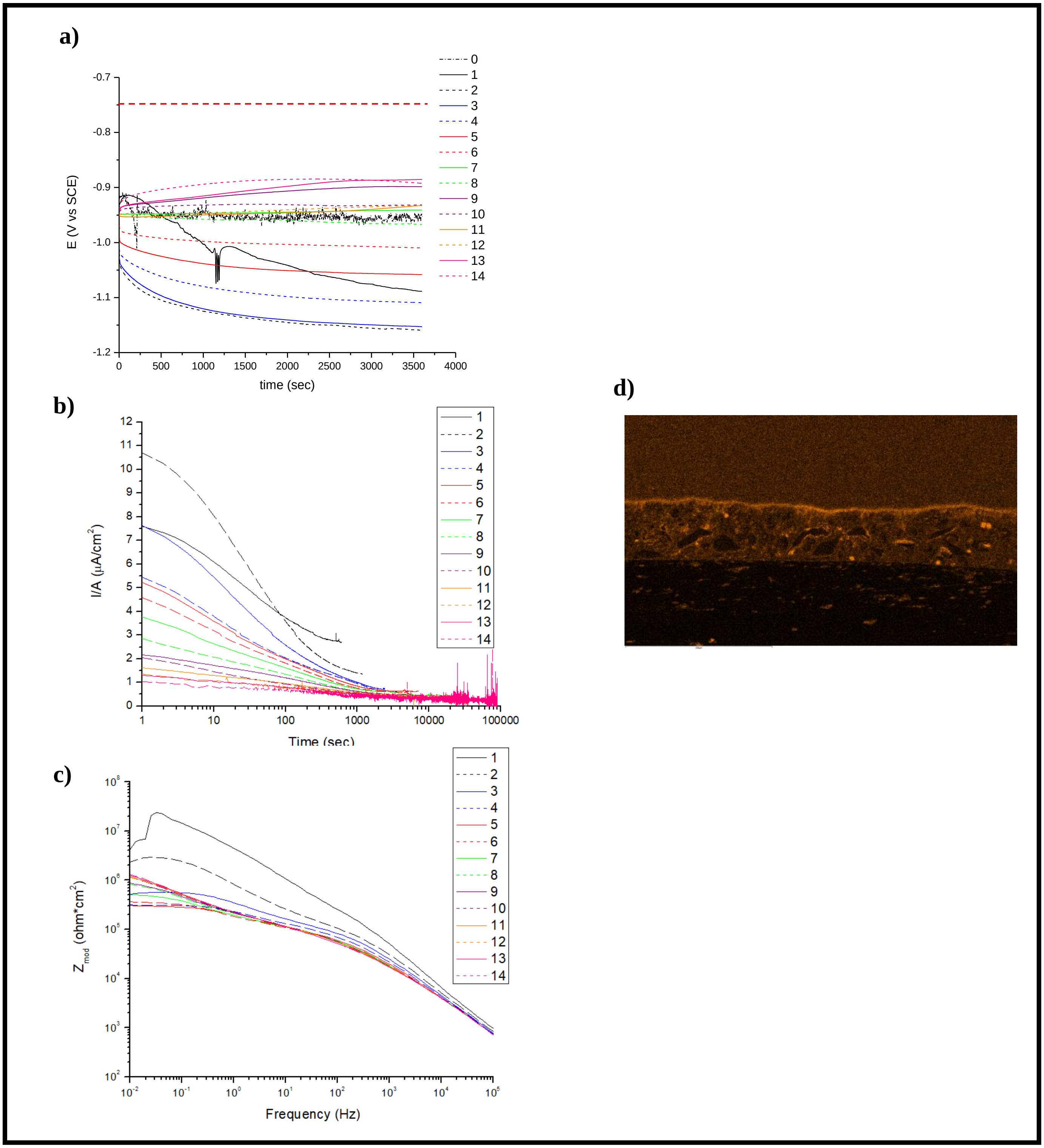

Figure 12. OCP/AC/DC cycle testing of MgAlRP applied to AA7075-T651 with DC potential hold at -0.95 $V_{SCE}$ and the legend denoting the cycle. **a)** Open circuit potential is shown with a red dotted line denoting the OCP of AA7075-T651, **b)** the current density output, **c)** residual

barrier properties shown in the Bode impedance response, and **d)** the oxygen EDS cross-section of MgRP after 100 net hours of DC potential hold at $-0.95V_{SCE}$. Testing is conducted under full immersion conditions in quiescent 0.6 M NaCl.

**Figure 13**

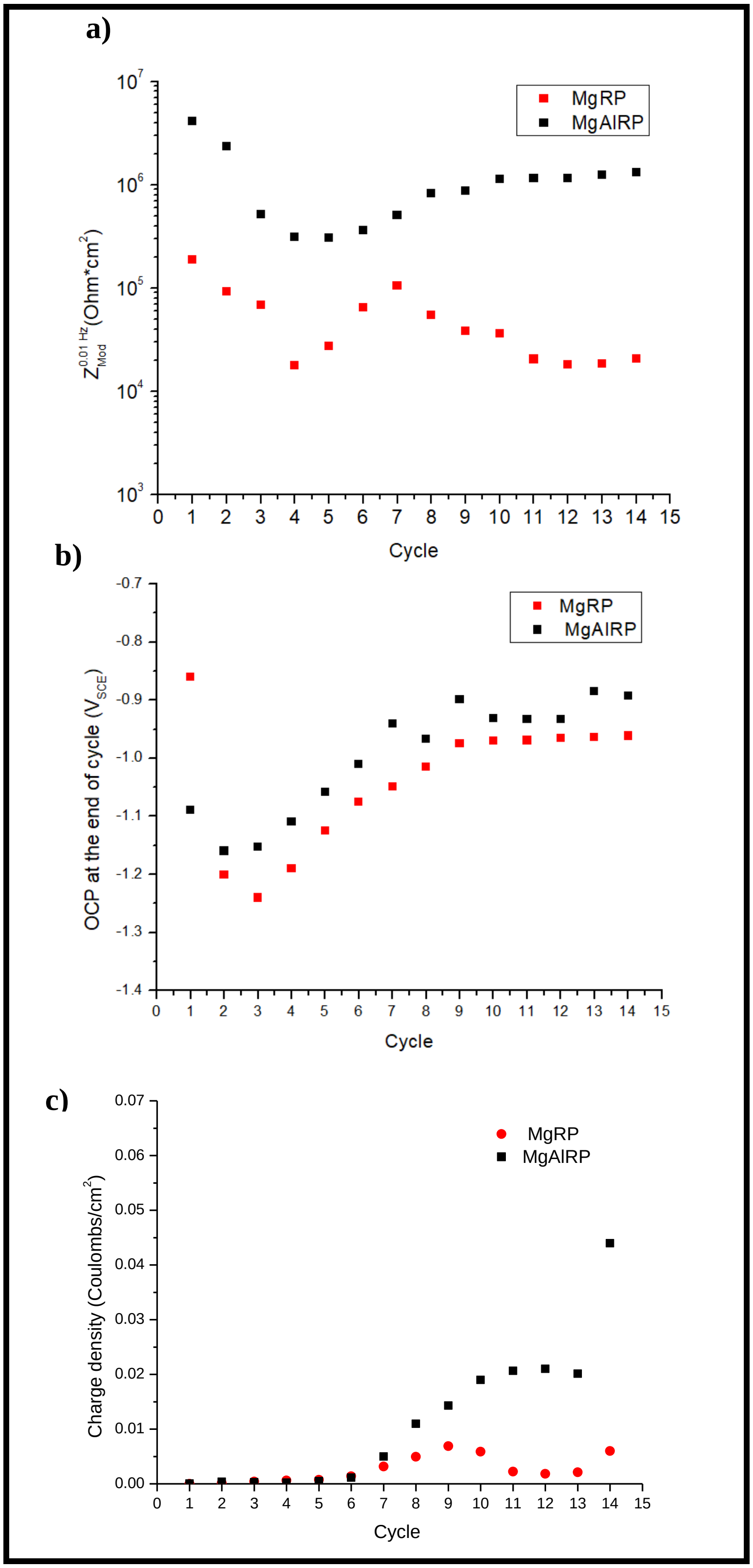
a)
MgRP
MgAlRP
$Z_{Mod}^{0.01 Hz}$ (Ohm*cm$^2$)
10^7
10^6
10^5
10^4
10^3
0 1 2 3 4 5 6 7 8 9 10 11 12 13 14 15
Cycle
b)
MgRP
MgAlRP
OCP at the end of cycle ($V_{SCE}$)
-0.7
-0.8
-0.9
-1.0
-1.1
-1.2
-1.3
-1.4
0 1 2 3 4 5 6 7 8 9 10 11 12 13 14 15
Cycle
c)
MgRP
MgAlRP
Charge density (Coulombs/cm$^2$)
0.07
0.06
0.05
0.04
0.03
0.02
0.01
0.00
0 1 2 3 4 5 6 7 8 9 10 11 12 13 14 15
Cycle

Figure 13. **a)** Variation of low frequency limit (0.01Hz) of $Z_{mod}$ against each potentiostatic cycle, **b)** the end of each OCP step shown against each cycle, and **c)** the charge density, as calculated from the current density, given for each DC/AC/OCP cycle.

**Figure 14**

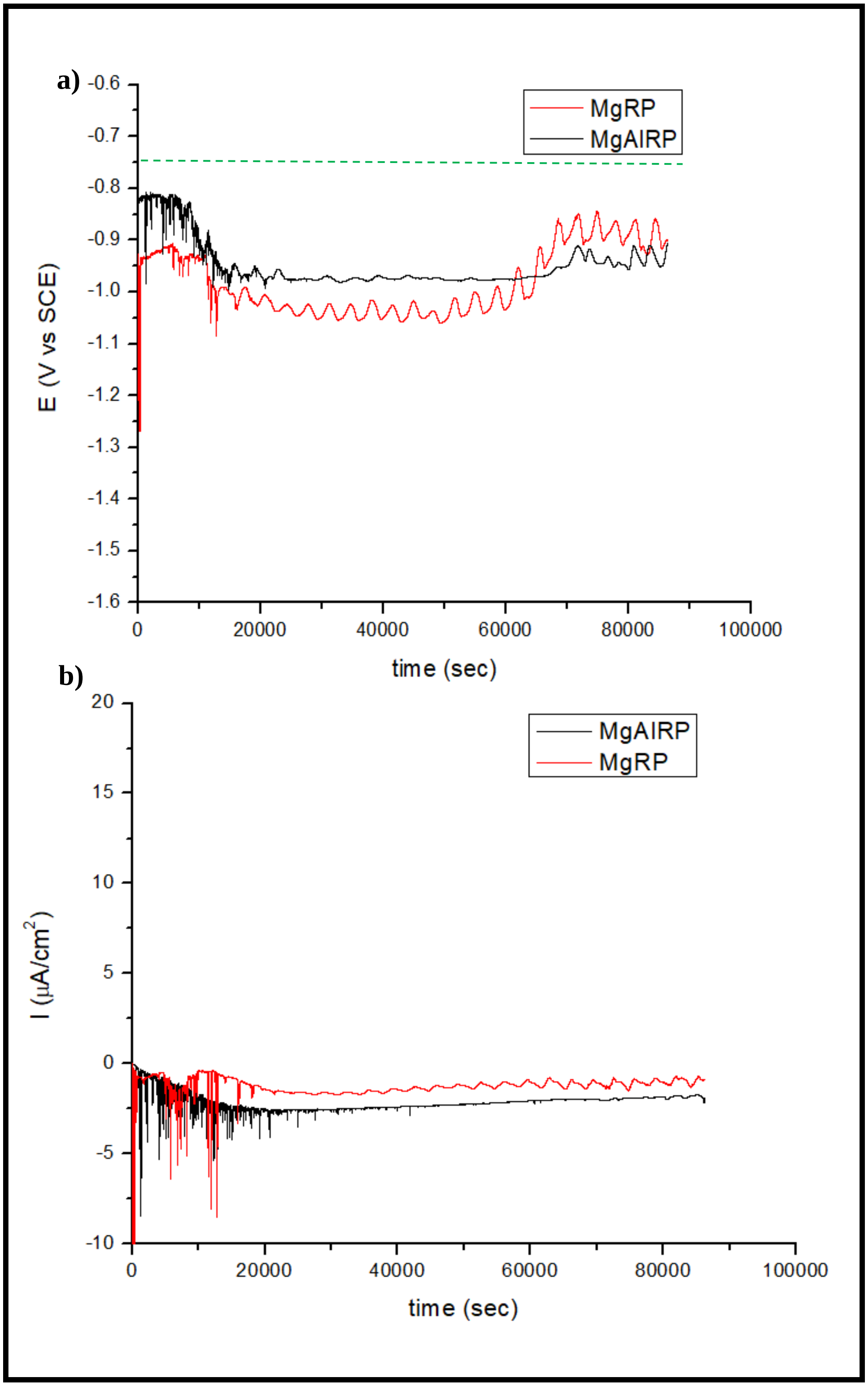

Figure 14. Galvanic corrosion of the coupled MgRP and MgAlRP coated AA7075-T651 to bare AA7075-T651 tested in quiescent 0.6 M NaCl in a 1:1 (MRP:AA 7075-T651) area ratio with the **a)** coupled potentials and **b)** coupled current densities. The green dashed line represents the OCP of bare AA7075-T651 (-0.75$V_{SCE}$) in quiescent 0.6 M NaCl.

**Figure 15**

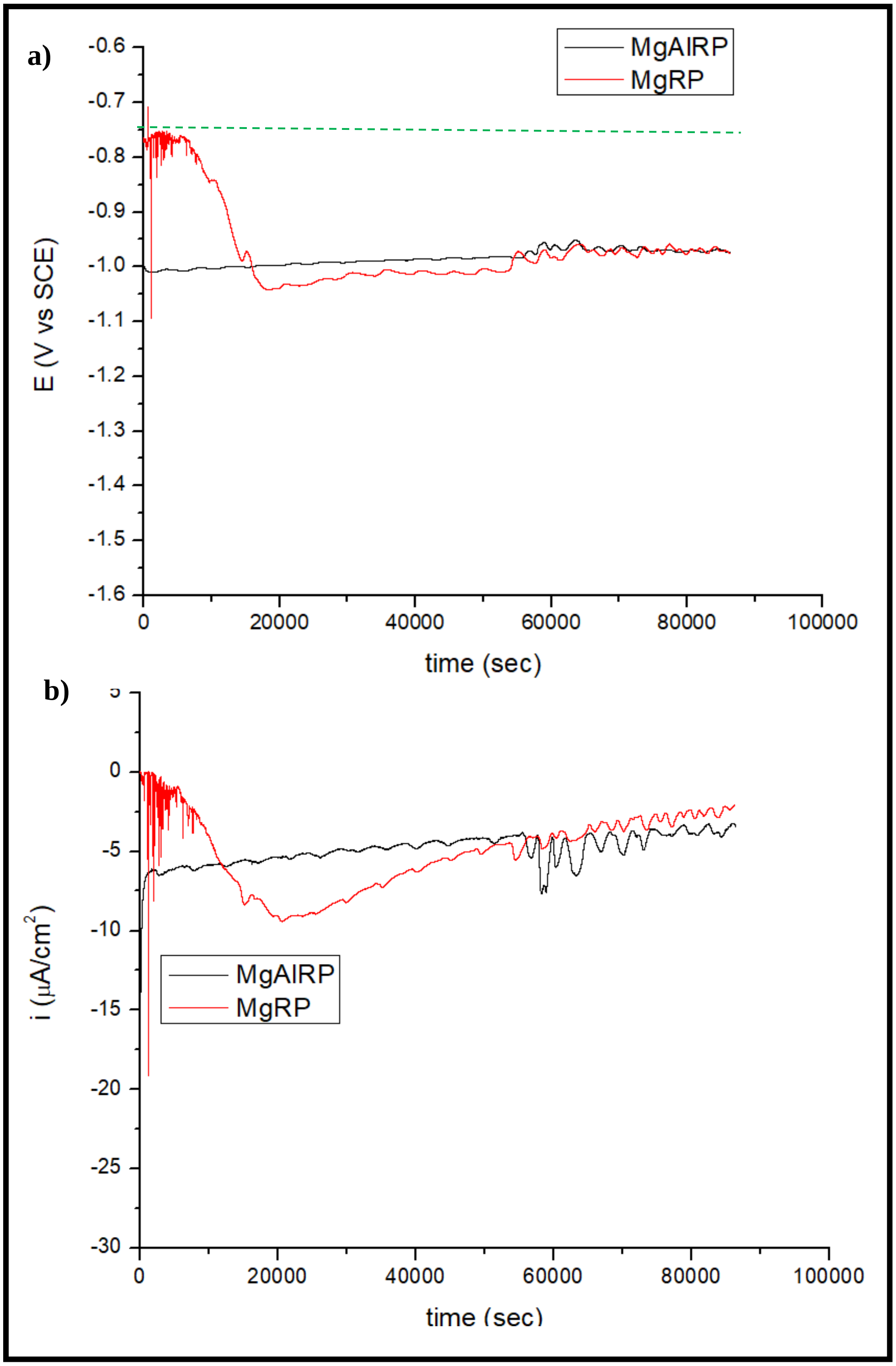

Figure 15. The galvanic corrosion testing of the coupled MgRP and MgAlRP coated AA7075-T651 to bare AA7075-T651 tested in quiescent 0.6 M NaCl in a 15:1 (MRP:AA 7075-T651) area ratio with the **a)** coupled potentials and **b)** coupled current densities. The green dashed line represents the OCP of bare AA7075-T651 (-0.75$V_{SCE}$) in quiescent 0.6 M NaCl.

**Figure 16**

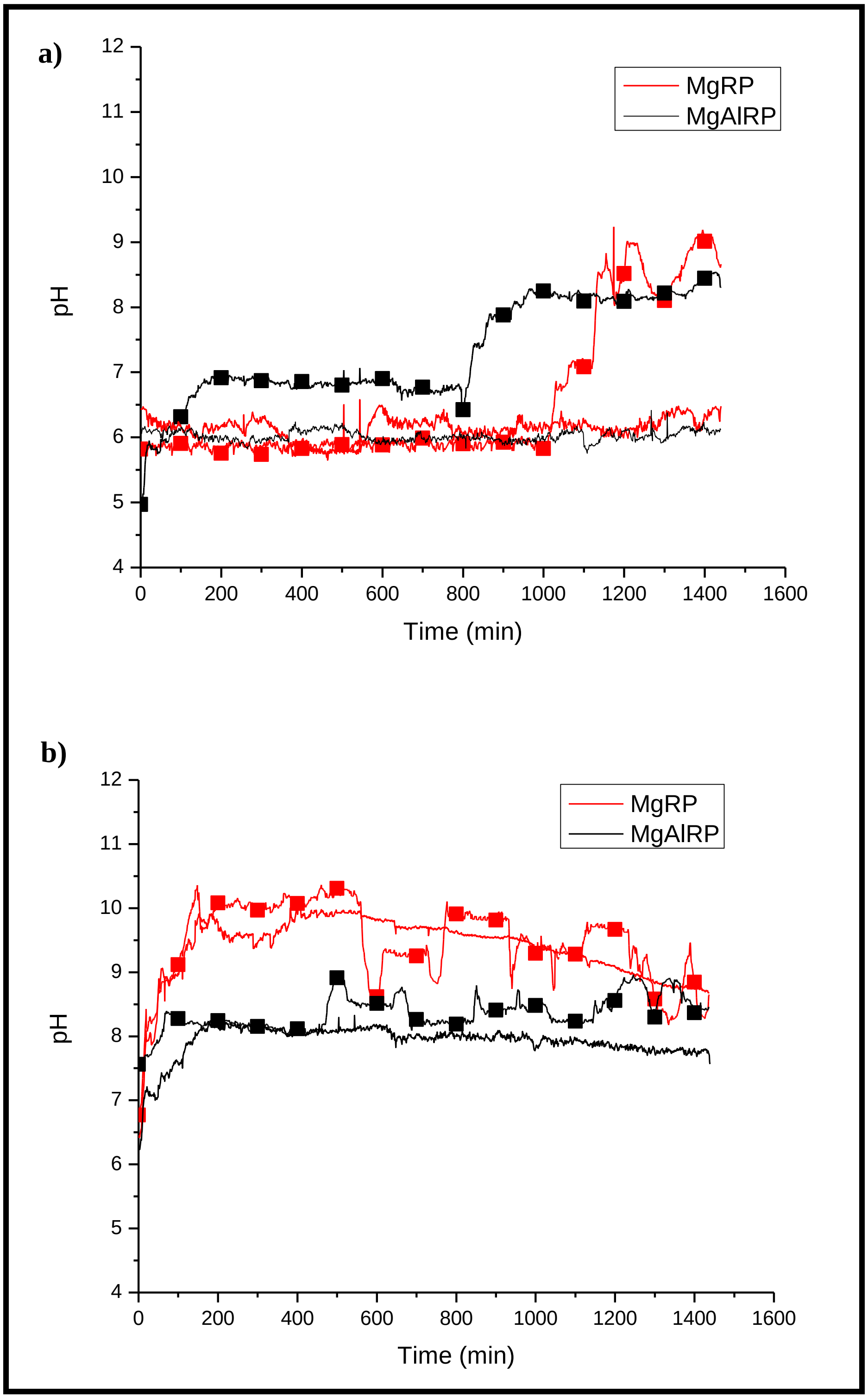

**Figure 16.** Local pH modification monitored throughout the galvanic coupling of each MRP – 7075-T651 exposure tested in 0.6M NaCl shown in **a)** over the bare AA7075-T651 WE and **b)**

over the MRP coating CE. The lines with square symbols denote the 15:1 (MRP:bare 7075) area ratio while the lines without symbols denote the 1:1 (MRP:bare 7075) area ratio.

**Figure 17**

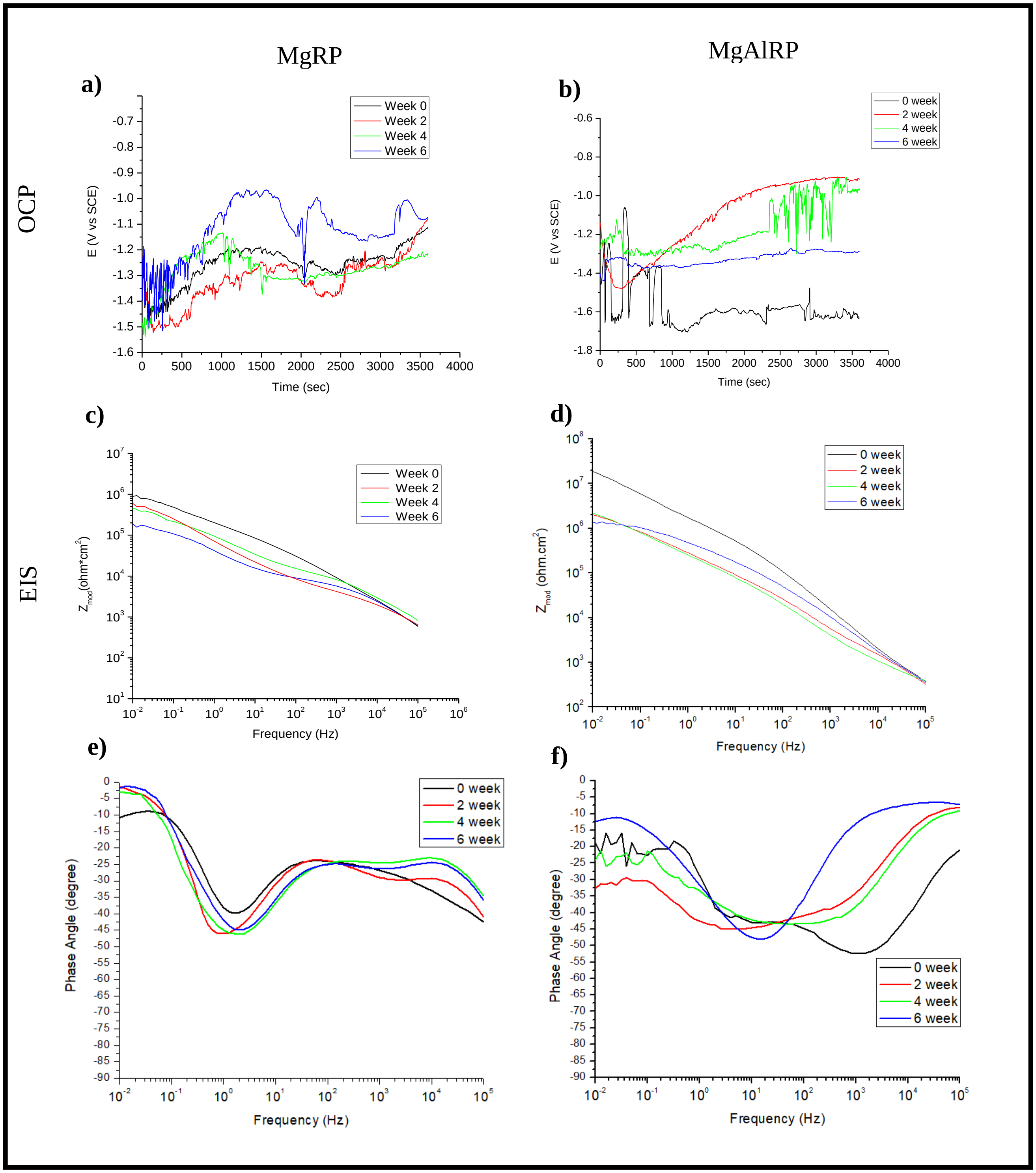

Figure 17. The open circuit potential and electrochemical impedance spectroscopy shown for both intact MgRP and MgAlRP coated AA7075-T651 throughout the six-week ASTM B117 accelerated environmental exposure testing in 0.6 M NaCl. Long term open circuit potential

shown for **a)** MgRP, and **b)** MgAlRP; the Bode magnitude is plotted for **c)** for MgRP, and **d)** MgAlRP. The phase angle progression is illustrated for **e)** MgRP, and **f)** MgAlRP.

**Figure 18**

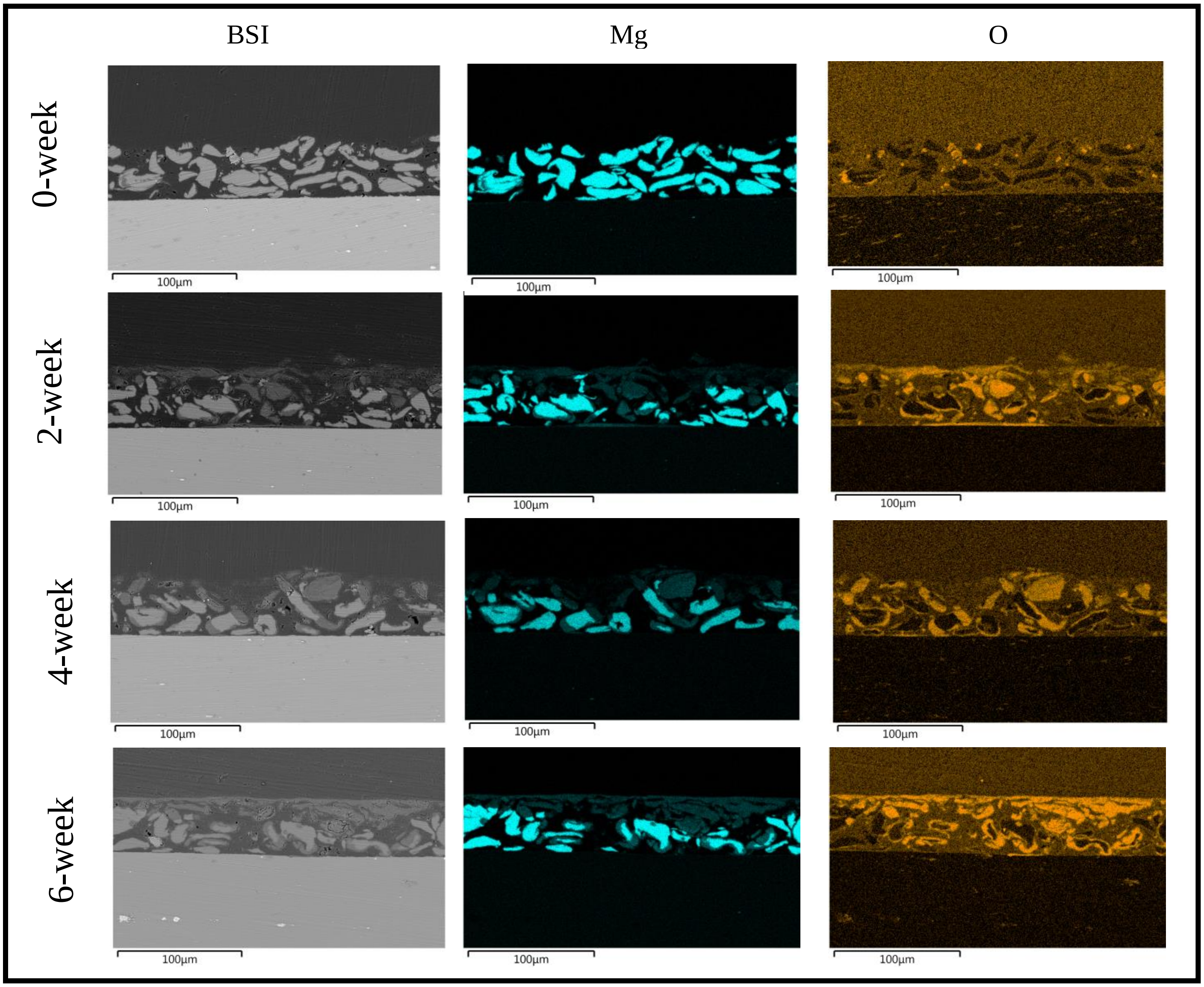

Figure 18. ASTM B117 salt spray testing on intact MgRP applied to AA7075-T651 across 0, 2 , 4, and 6 weeks of accelerated environmental exposure in 0.6M NaCl as shown in BSI SEM micrographs with individual magnesium and oxygen signals from EDS elemental mapping.

**Figure 19**

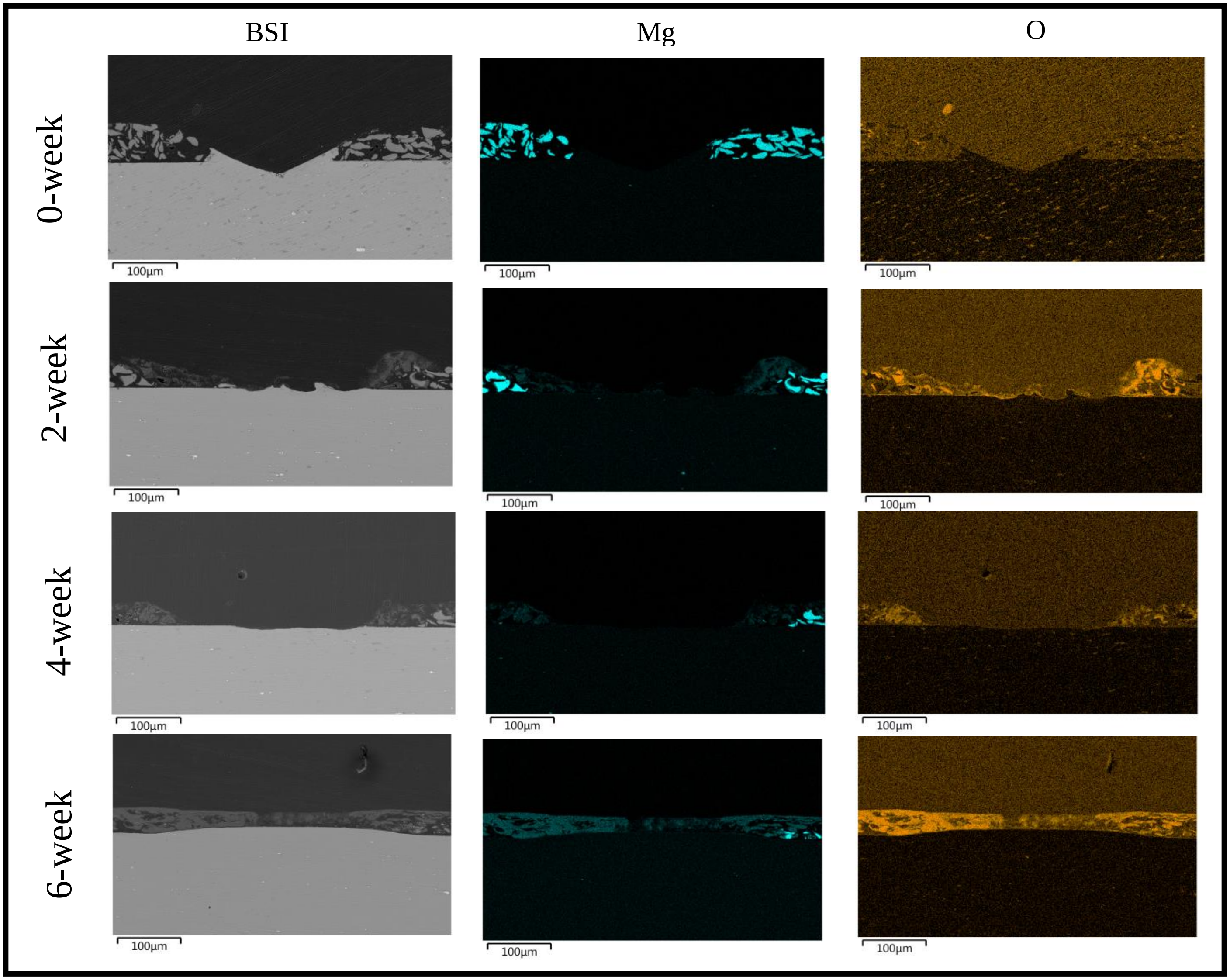

Figure 19. ASTM B117 salt spray testing on scribed MgRP applied to AA7075-T651 across 0, 2 , 4, and 6 weeks of accelerated environmental exposure in 0.6M NaCl as shown in BSI SEM micrographs with individual magnesium and oxygen signals from EDS elemental mapping.

**Figure 20**

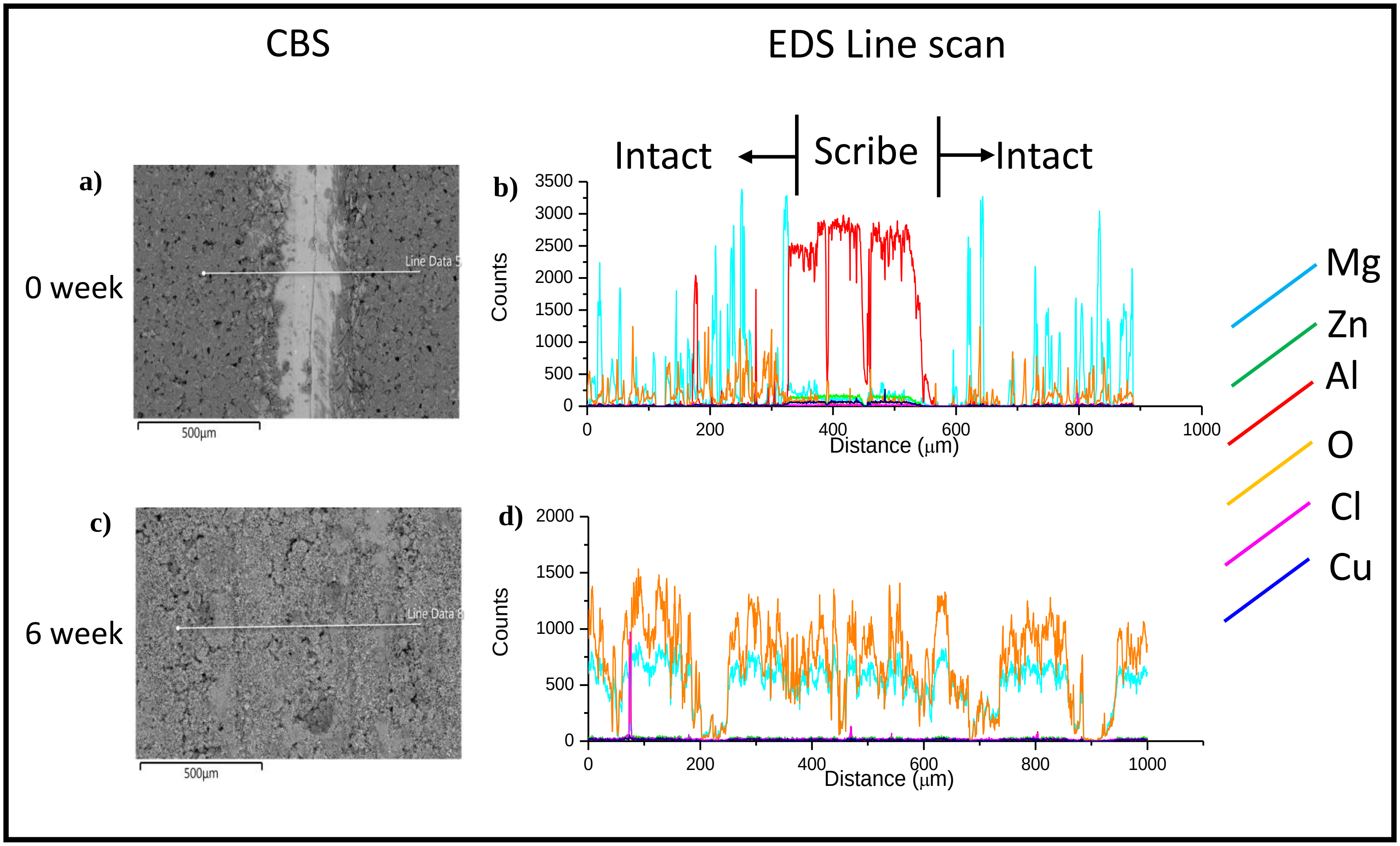

Figure 20. ASTM B117 salt spray testing of MgRP shown in plan-view BSI SEM micrographs for zero and six-week in **a)** and **c)** with EDS line scans shown in **b)** and **d)**.

**Figure 21**

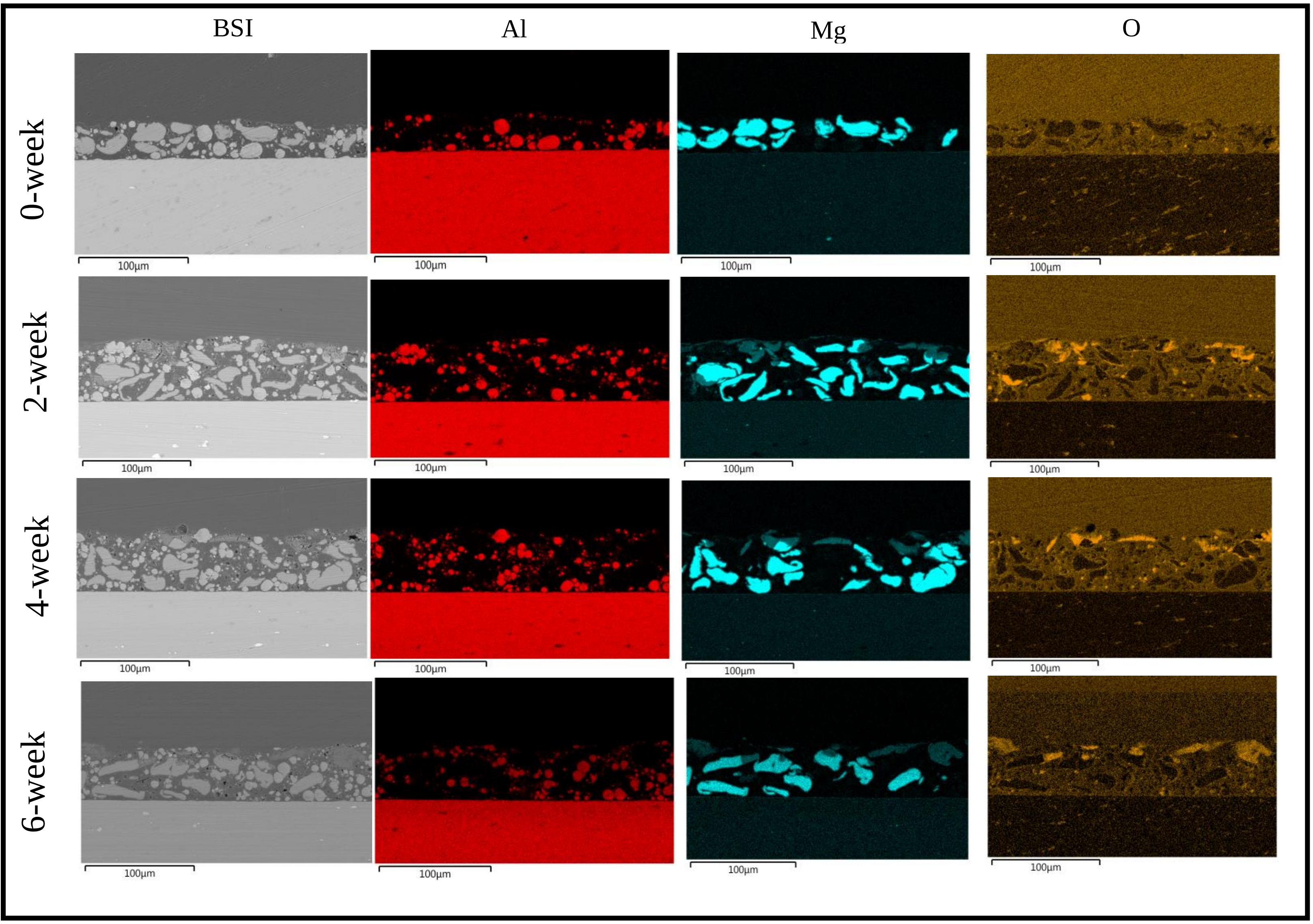

Figure 21. ASTM B117 salt spray testing on intact MgAlRP applied to AA7075-T651 across 0, 2 , 4, and 6 weeks of accelerated environmental exposure in 0.6M NaCl as shown in BSI SEM micrographs with individual magnesium and oxygen signals from EDS elemental mapping.

**Figure 22**

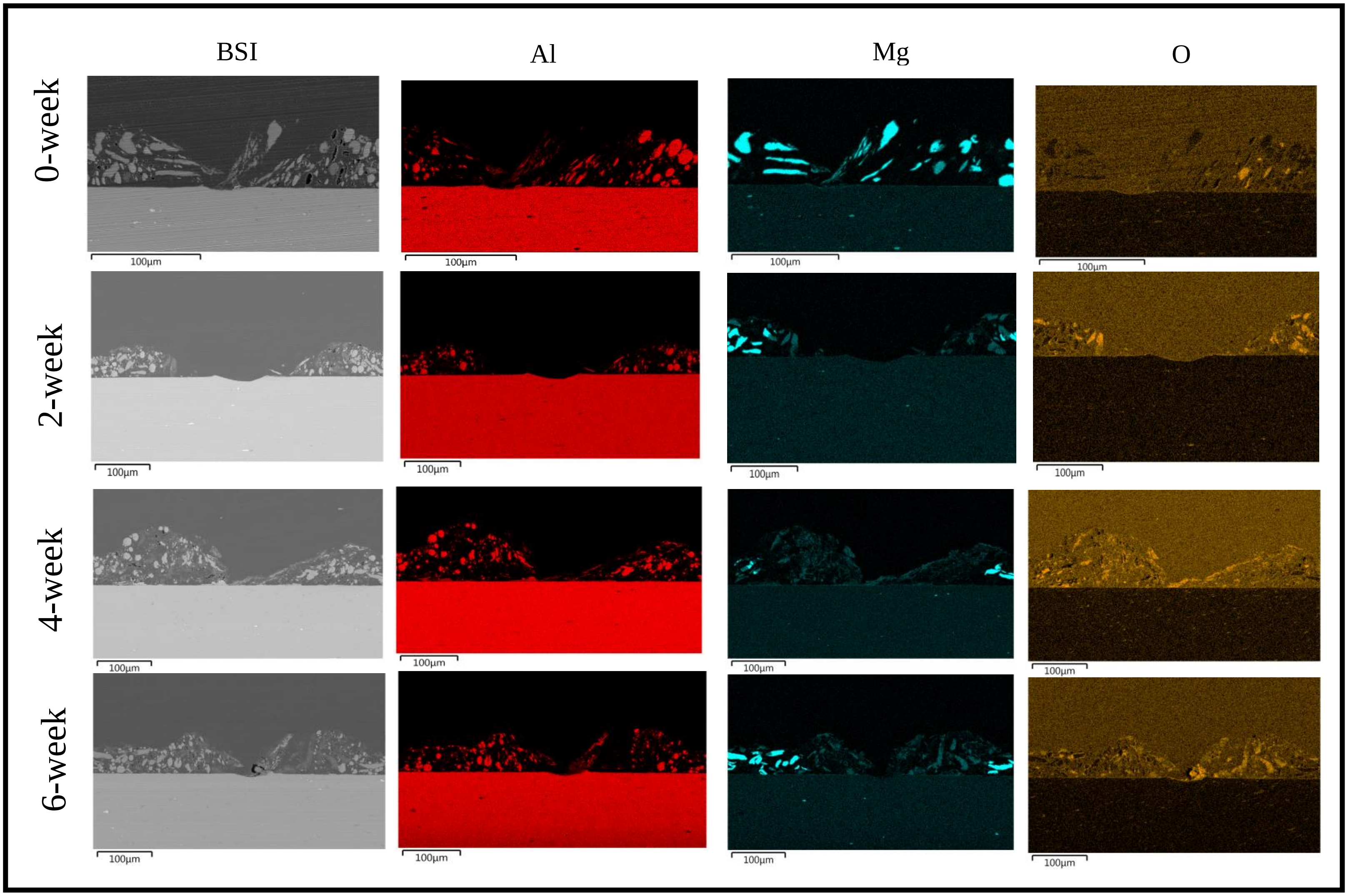

Figure 22. ASTM B117 salt spray testing on scribed MgAlRP applied to AA7075-T651 across 0, 2 , 4, and 6 weeks of accelerated environmental exposure in 0.6M NaCl as shown in BSI SEM micrographs with individual magnesium and oxygen signals from EDS elemental mapping.

**Figure 23**

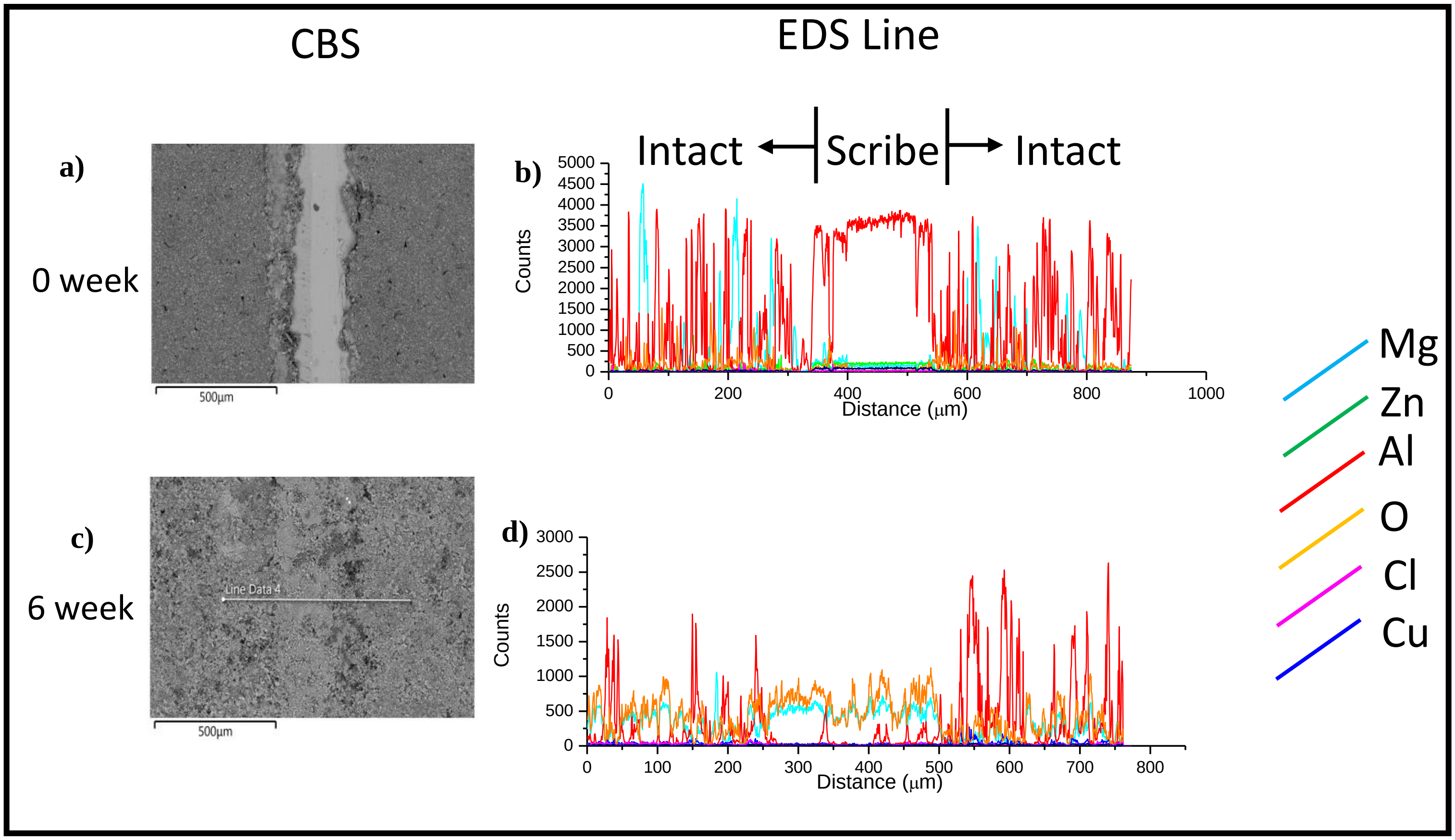

Figure 23. ASTM B117 salt spray testing of MgAlRP shown in plan view BSI SEM micrographs are shown for zero and six-week in **a)** and **c)** with EDS line scans shown in **b)** and **d)**.

**Figure 24**

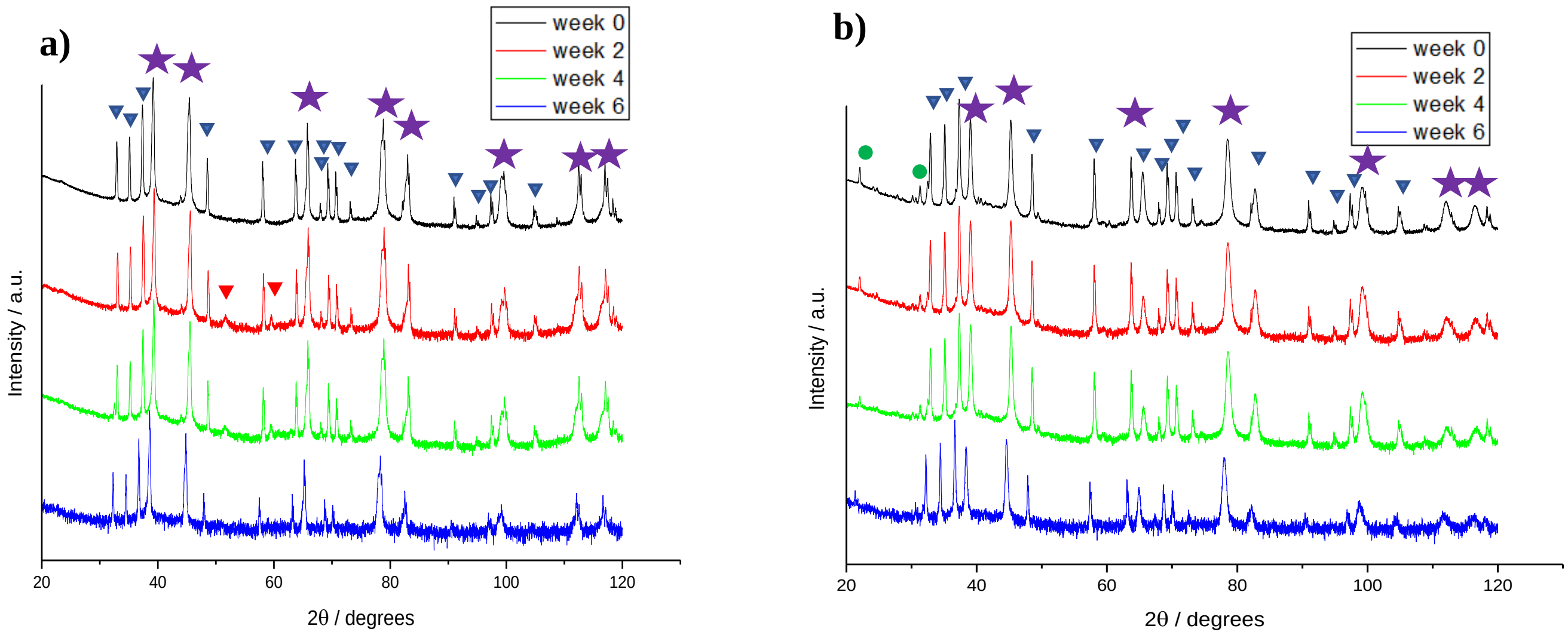

Figure 24. X-ray diffraction analysis on **a)** intact MgRP and **b)** intact MgAlRP condition following ASTM B117 exposure in 0.6M NaCl. In the XRD spectra purple stars denotes FCC Al peaks, blue triangles denote Mg peaks, red triangles denote $Mg(OH)_2$, and green circles denote $Al(OH)_3$.

**Figure 25**

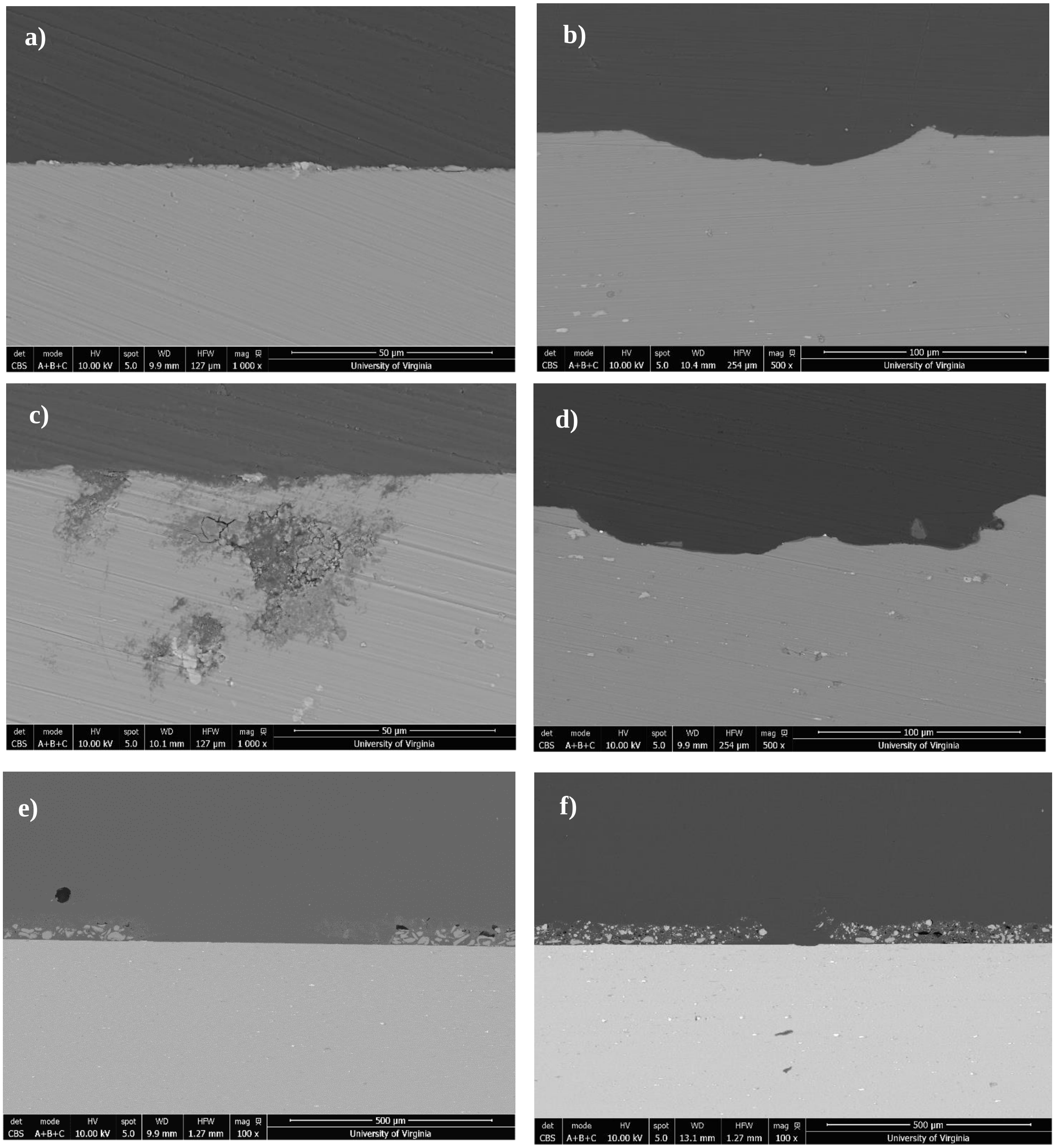

Figure 25. Nitric-washed pristine bare AA 7075-T651 control BSI with no protection scheme shown over the uncoated bare region in **a)** and over the scribe shown in **b)**. Nitric washed post six-week bare AA 7075-T651 ASTM B117 BSI with no protection scheme is shown over the bare region in **c)** and scribed region **d)**. The post six-week MgRP coated AA7075-T651 ASTM

B117 salt spray testing over the scribed region is shown in **e)**. The post six-week MgAlRP coated AA7075-T651 ASTM B117 salt spray testing over the scribed region is shown in **f)**.

**Figure 26**

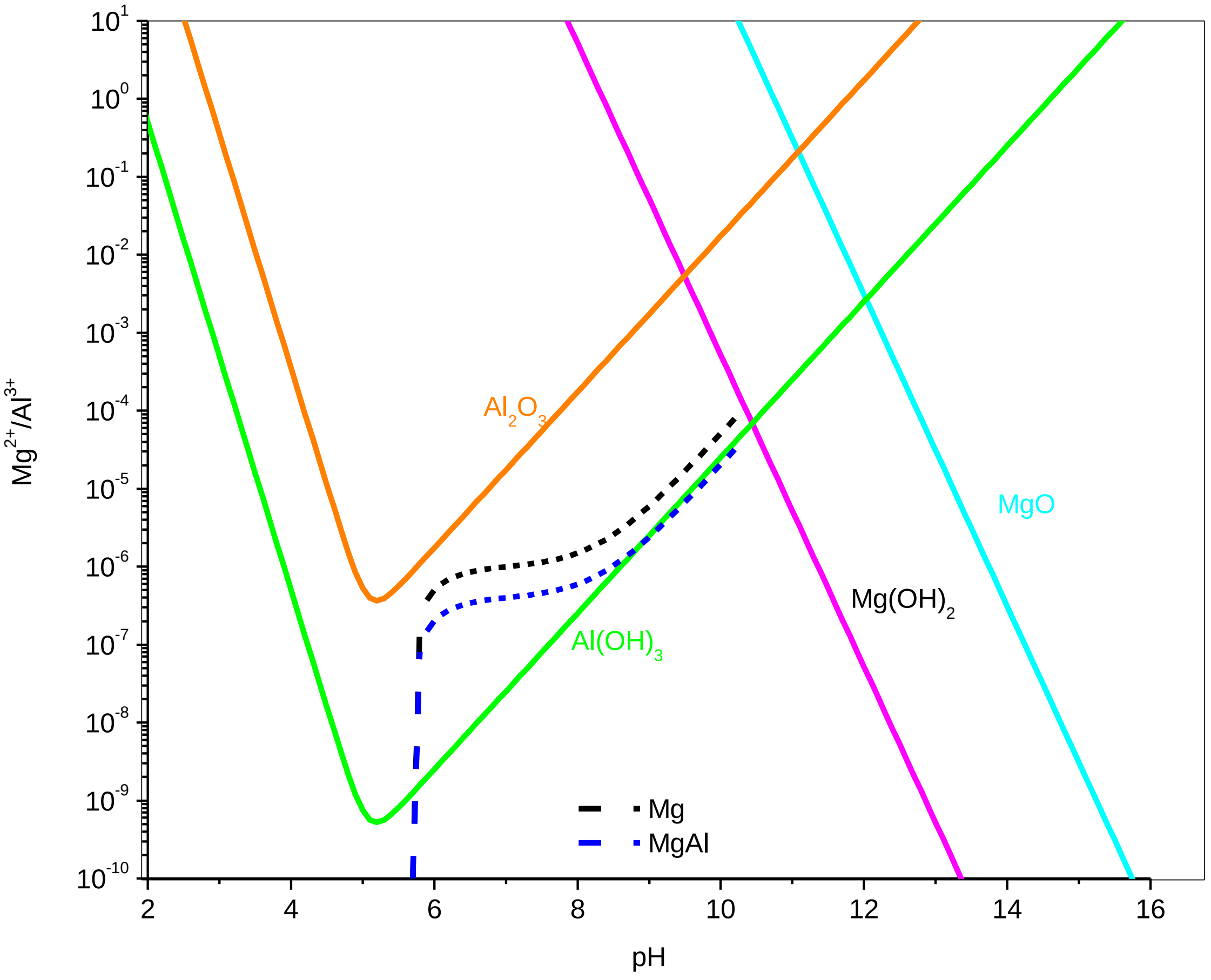

Figure 26. Chemical stability diagram depicting the chemical equilibria lines of $Al^{3+}/Al_2O_3$, $Al^{3+}/Al(OH)_3$, $Mg^{2+}/MgO$, $Mg^{2+}/Mg(OH)_2$ using solid lines. The dissolution trajectory of the corroding Mg (dashed black line) and MgAlRP (dashed blue line) systems is dependent on the initial solution chemistry.

**Figure 27**

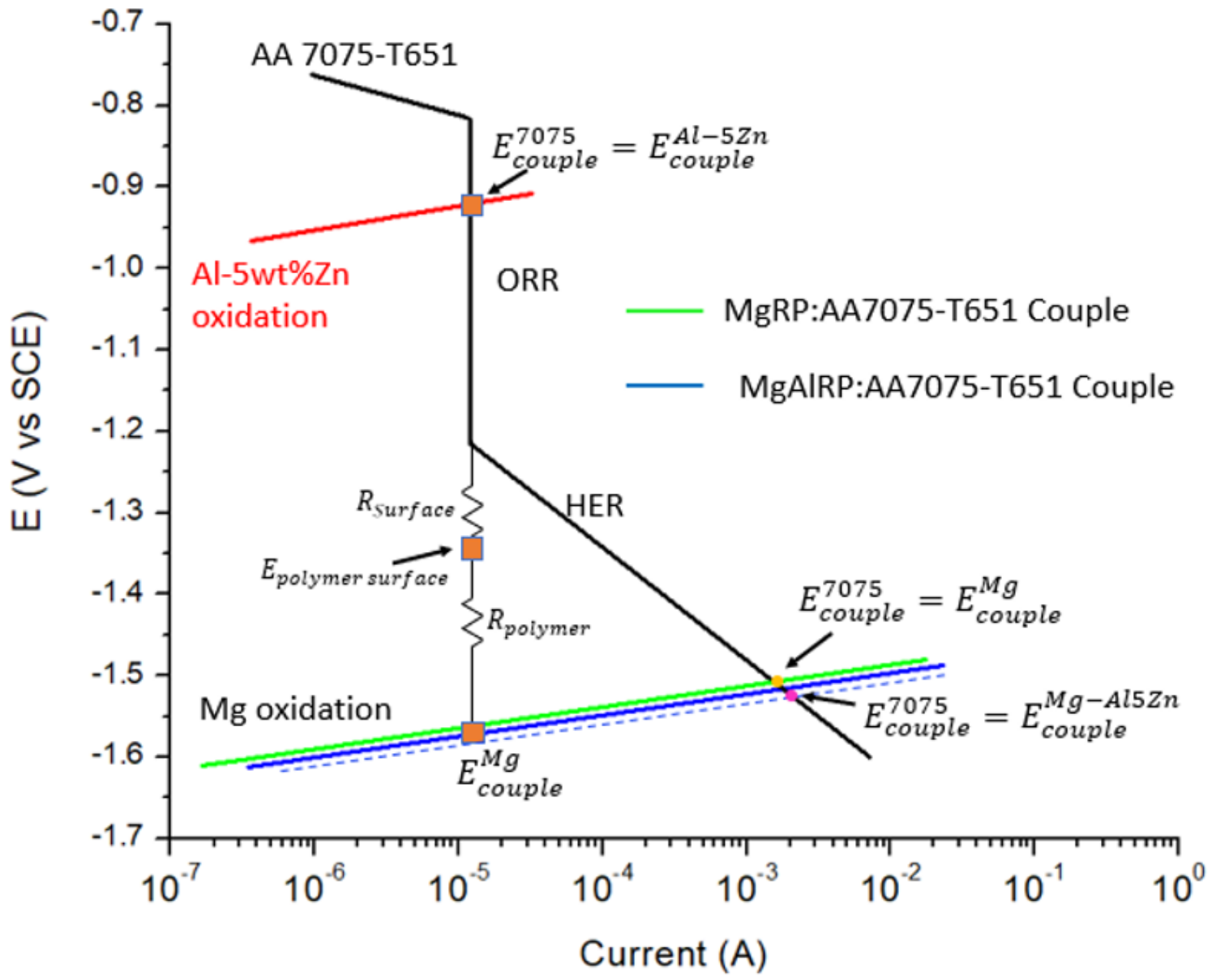

Figure 27. Schematic representation of the galvanic couple between formed in both MgRP and MgAlRP coated AA 7075-T651 substrate exposed to unadjusted quiescent 0.6 M NaCl.